\documentclass{aastex701}

\usepackage{graphicx} 
\usepackage{amsmath}
\usepackage{braket}
\usepackage{amssymb}

\usepackage{relsize}
\usepackage{listings}
\definecolor{dkgreen}{rgb}{0,0.6,0}
\definecolor{gray}{rgb}{0.5,0.5,0.5}
\definecolor{mauve}{rgb}{0.58,0,0.82}
\def\bm#1{\bs{#1}}

\def\gpcoh{\,h^{-1}{\rm Gpc}}
\def\hompc{\,h\,{\rm Mpc}^{-1}}

\newcommand{\bs}[1]{\boldsymbol{#1}}
\newcommand{\dd}{\mathrm{d}}

\begin{document}
\title{Efficient estimators for power spectrum and bispectrum multipole measurements}

\correspondingauthor{Gong-Bo Zhao, \ Florian Beutler \ \& John Peacock}
\email{gbzhao@nao.cas.cn, \ florian.beutler@ed.ac.uk, \ jap@roe.ac.uk}

\author[0009-0000-6795-0922]{Yunchen Xie}
\email{Y.Xie-102@sms.ed.ac.uk}
\affiliation{National Astronomical Observatories, Chinese Academy of Sciences, Beijing, 100101, P.R.China}
\affiliation{University of Chinese Academy of Sciences, Beijing, 100049, P.R.China}
\affiliation{Institute for Astronomy, University of Edinburgh, Royal Observatory, Blackford Hill, Edinburgh EH9 3HJ, UK}

\author[0000-0002-7284-7265]{Ruiyang Zhao}
\email{zhaoruiyang19@mails.ucas.edu.cn}
\affiliation{National Astronomical Observatories, Chinese Academy of Sciences, Beijing, 100101, P.R.China}
\affiliation{University of Chinese Academy of Sciences, Beijing, 100049, P.R.China}
\affiliation{Institute of Cosmology \& Gravitation, University of Portsmouth, Dennis Sciama Building, Portsmouth, PO1 3FX, UK}
\affiliation{Department of Astronomy, Tsinghua University, Beijing 100084, China}

\author[0009-0007-9215-489X]{Gan Gu}
\email{gugan20@mails.ucas.ac.cn}
\affiliation{National Astronomical Observatories, Chinese Academy of Sciences, Beijing, 100101, P.R.China}
\affiliation{University of Chinese Academy of Sciences, Beijing, 100049, P.R.China}

\author[0000-0003-0216-1230]{Xiaoma Wang}
\email{wangxiaoma20@mails.ucas.ac.cn}
\affiliation{National Astronomical Observatories, Chinese Academy of Sciences, Beijing, 100101, P.R.China}
\affiliation{University of Chinese Academy of Sciences, Beijing, 100049, P.R.China}

\author[0000-0002-2671-9078]{Xiaoyong Mu}
\email{mouxiaoyong15@mails.ucas.ac.cn}
\affiliation{National Astronomical Observatories, Chinese Academy of Sciences, Beijing, 100101, P.R.China}
\affiliation{University of Chinese Academy of Sciences, Beijing, 100049, P.R.China}

\author[0000-0001-7756-8479]{Yuting Wang}
\email{ytwang@nao.cas.cn}
\affiliation{National Astronomical Observatories, Chinese Academy of Sciences, Beijing, 100101, P.R.China}

\author[0000-0003-4726-6714]{Gong-Bo Zhao}
\email{gbzhao@nao.cas.cn}
\affiliation{National Astronomical Observatories, Chinese Academy of Sciences, Beijing, 100101, P.R.China}
\affiliation{University of Chinese Academy of Sciences, Beijing, 100049, P.R.China}

\author[0000-0003-0467-5438]{Florian Beutler}
\email{florian.beutler@ed.ac.uk}
\affiliation{Institute for Astronomy, University of Edinburgh, Royal Observatory, Blackford Hill, Edinburgh EH9 3HJ, UK}

\author[0000-0002-1168-8299]{John A. Peacock}
\email{jap@roe.ac.uk}
\affiliation{Institute for Astronomy, University of Edinburgh, Royal Observatory, Blackford Hill, Edinburgh EH9 3HJ, UK}

\begin{abstract}
Large galaxy surveys demand fast and scalable estimators for anisotropic clustering statistics beyond the monopole.
We present a suite of efficient FFT-based estimators for power-spectrum and bispectrum multipoles, built upon exact
conjugation and parity symmetries of spherical-harmonic--weighted Fourier transforms of real fields. These symmetries eliminate redundant magnetic sub-configurations, thereby reducing the computational cost by a factor of 2. For the Yamamoto power-spectrum multipoles, we further decrease the cost of high-order even multipoles by algebraically expressing $\mathcal{L}_{2n}$ in terms of lower-order
Legendre polynomials, thereby measuring modified high-order multipoles using only low-$\ell$ fields with a small and
controlled deviation from the traditional definition. We introduce a new TripoSH bispectrum
estimator obtained by compressing the Scoccimarro bispectrum along an alternative triangle side, which substantially
reduces the FFT scaling for commonly used quadrupole configurations in the large-$k$-bin limit.
We also derive an analytic treatment of bispectrum shot noise by integrating spherical-harmonic kernels over the
triangle-constrained $k$-space volumes, avoiding additional FFTs or costly spherical-Bessel evaluations and enabling
fast and accurate shot-noise subtraction. Based on these optimizations, we also introduce \href{https://github.com/YunchenXie/CosmoNPC}{\texttt{CosmoNPC}}, an open-source Python package for large-scale-structure clustering measurements.

\end{abstract}


\section{Introduction} \label{sec:intro}

Galaxy redshift surveys map the late-time matter distribution over cosmological volumes, enabling precision tests of the standard cosmological model and its extensions. A primary summary statistic is the galaxy power spectrum, whose anisotropies with respect to the line of sight (LOS)
encode both redshift-space distortions (RSD) and geometric distortions via the Alcock--Paczynski effect.
In practice, this anisotropy is commonly compressed into multipoles of $P(k,\mu)$, which have been extensively used in analyses of BOSS and eBOSS data to extract the BAO scale and the growth rate of structure
(e.g., \citealt{Beutler_2014,Gil_Mar_n_2020}; see also \citealt{Wang_2024} for recent discussions of information content in
power-spectrum multipoles).
Beyond two-point statistics, three-point functions such as the bispectrum provide access to additional non-Gaussian
information generated by non-linear gravitational evolution and galaxy bias
\citep{1998ApJ...496..586S,Scoccimarro_2000}, and have been shown to improve cosmological constraints when combined
with the power spectrum \citep[e.g.,][]{2015MNRAS.451..539G,2017MNRAS.465.1757G}.
This higher-order information is also of broad interest for constraining neutrino mass and primordial non-Gaussianity
\citep{Hahn_2020,Hahn_2021,heinrich2023measuring}.

A central challenge is that next-generation analyses demand not only accurate estimators, but also estimators that are
fast enough to be applied to thousands of mock catalogs for covariance estimation and systematic validation.
For the power spectrum, the optimal quadratic estimator in the plane-parallel limit is the classic FKP estimator
\citep{Feldman:1993ky}.
However, realistic surveys require a varying LOS across the footprint, motivating ``moving-LOS'' estimators such as
the Yamamoto estimator for power-spectrum multipoles \citep{10.1093/pasj/58.1.93}.
Na\"ively, evaluating the LOS-dependent kernel $\mathcal{L}_\ell(\hat{\bm k}\!\cdot\!\hat{\bm r})$ would require an explicit sum
over all galaxy pairs and is therefore computationally prohibitive.
A major advance was the realization that these multipoles can be measured efficiently with Fast Fourier Transforms
(FFTs) by rewriting the LOS-dependent factors into separable angular pieces.
This has been achieved via Cartesian decompositions \citep{PhysRevD.92.083532,Bianchi_2015} and, more efficiently,
via spherical-harmonic decompositions \citep{Hand_2017}, where a given multipole of order $\ell$ can be obtained using
$2\ell+1$ spherical-harmonic--weighted FFTs rather than the $(\ell+1)(\ell+2)/2$ FFTs required by the Cartesian method.

The computational burden is even more acute for bispectrum measurements.
Fast bispectrum estimators based on FFTs were developed by \citet{PhysRevD.92.083532} and have since been applied to
survey data, demonstrating the value of joint power-spectrum--bispectrum analyses \citep[e.g.,][]{2017MNRAS.465.1757G}.
More recently, \citet{2019MNRAS.484..364S} presented a complete FFT-based formalism for decomposing the redshift-space
bispectrum into a tri-polar spherical harmonic (TripoSH) basis, providing a systematic way to capture anisotropy in
three-point clustering.
Such decompositions build on the general theory of isotropic $N$-point basis functions \citep{cahn2020isotropic} and
are closely connected to efficient algorithms for higher-point statistics \citep{2022PNAS..11911366P}.
Despite these advances, practical bispectrum multipole measurements remain expensive because (i) many angular
sub-configurations must be evaluated and binned, (ii) the triangle closure in Fourier space must be performed for many
$(k_1,k_2,k_3)$ configurations, and (iii) the corresponding shot-noise terms are non-trivial and can themselves require
additional costly operations \citep{2019MNRAS.484..364S,PhysRevD.92.083532}.

In this paper, we develop a set of optimizations and new estimators that substantially reduce the computational cost
of power-spectrum and bispectrum multipole measurements while preserving the target information content.
Our approach is based on a general symmetry of spherical-harmonic--weighted Fourier transforms of real fields, which
relates magnetic sub-components $(\ell,m)$ to $(\ell,-m)$ under conjugation and parity.
This symmetry implies that many of the FFTs traditionally computed in multipole estimators are redundant.
We exploit this observation at three levels:
(i) we provide a \emph{lossless} reduction of the number of FFTs required by existing spherical-harmonic estimators for
the Yamamoto power-spectrum multipoles and for FFT-based bispectrum estimators, eliminating redundant
$m<0$ components and reducing the FFT scaling by nearly a factor of two at fixed $\ell$;
(ii) we further reduce the cost of high-order even power-spectrum multipoles by using algebraic identities among
Legendre polynomials to express $\mathcal{L}_{2n}$ in terms of lower-order polynomials, enabling ``modified'' high-$\ell$ multipole
measurements with only low-$\ell$ fields and a controlled deviation from the traditional definition;
and (iii) we introduce a new TripoSH bispectrum estimator obtained by compressing the Scoccimarro bispectrum along
an alternative triangle side, which reduces the FFT scaling for commonly used quadrupole configurations in the limit
of many $k$-bins.

We validate these methods using standard large-volume mock catalogs designed to reproduce BOSS-like selections,
including MultiDark--Patchy mocks \citep{10.1093/mnras/stv2826,10.1093/mnras/stw1014,pub.1038954845}, and we assess
information content for bispectrum configurations using the Molino/Quijote-based suites
\citep{Villaescusa_Navarro_2020,Hahn_2021}.
Finally, we address a key practical bottleneck: shot-noise subtraction for bispectrum multipoles.
We show that, for shot-noise terms, the angular structure of the triangle-constrained integrals can be treated
analytically, allowing accurate shot-noise evaluation without introducing extra FFTs or requiring real-time spherical
Bessel evaluations as in the standard Rayleigh-expansion approach \citep{2019MNRAS.484..364S}.

The remainder of the paper is organized as follows.
In \S\ref{sec:level1} we present the symmetry-based, lossless optimizations to existing power-spectrum and bispectrum
multipole estimators.
In \S\ref{sec:level2} we describe the Legendre-polynomial reduction strategy for efficient high-order even multipoles.
In \S\ref{sec:level3} we introduce the new compressed TripoSH bispectrum estimator and compare its scaling and
information content to existing approaches.
In \S\ref{sec:shotnoise} we derive the analytical treatment of bispectrum shot noise and validate it against FFT- and
Rayleigh-based computations.
In \S\ref{sec:cosmonpc} we introduce \texttt{CosmoNPC}\footnote{The source code is publicly available at \url{https://github.com/YunchenXie/CosmoNPC}}.
, a user-friendly open-source Python library that implements the estimators developed in this work.
We conclude in \S\ref{sec:conclusions}.

\section{Lossless Symmetry Reduction of Existing Estimators}\label{sec:level1}

The spherical harmonics weighted Fourier transform,
\begin{equation}\label{eq:key1}
    S_\ell^m(\bs{k}) = \int \dd^3x\,S(\bs{x})Y_\ell^m(\hat{\bs{x}}){\mathrm e}^{-{\rm i}\bs{k}\cdot\bs{x}}\,,
\end{equation}
is heavily used in power spectrum and bispectrum estimation, where $Y_\ell^m$ follows the Condon–Shortley phase convention. If $S(\bs{x})$ is a real-valued function, negative integer $m$-components can be related to positive components by the following symmetry relation:
\begin{equation}
    S_\ell^{-m}(\bs{k}) = (-1)^m S_\ell^{m*}(-\bs{k})\,,
\end{equation}
where $*$ denotes the complex conjugate. One can find the proof of this property in Appendix~\ref{sec:proofs}. This is the core formula of this work, and we will use this relation repeatedly throughout the paper.

\subsection{Yamamoto estimator}\label{sec:l1_yama}

Power spectrum multipoles $\widehat{P}_{\ell}(k)$ can be estimated as follows \citep{10.1093/pasj/58.1.93}:
\begin{equation}\label{eq:Hand17}
    \widehat{P}_{\ell}(k)=\frac{2\ell+1}{I_{22}}\int \frac{\dd^2 \hat{k}}{4\pi}F_0(\bs{k})F_{\ell}^*(\bs{k})-N_{\ell}^{\rm shot}\,,
\end{equation}
where $\ell = 0,2,4,\dots$, the normalization factor $I_{22} \equiv \alpha \sum_{j=1}^{N_r} \bar{n}(\bs{x}_j) w_j^2$, $\alpha$ is the ratio of the weighted number of samples in a galaxy catalog $N_g$ to that of the random catalog $N_r$, $\bar{n}$ is the mean number density and $w$ denotes the weight. The shot noise is calculated as $N_{\ell}^{\rm shot}=\delta_{\ell,0}^{\rm K}I_{22}^{-1}(\sum_{j=1}^{N_g}+\alpha^2 \sum_{j=1}^{N_r}) w_j^2$. The quantity $F_{\ell}(\bs{k})$ is evaluated as
\begin{equation}\label{eq:F_ell}
    F_{\ell}(\bs{k})  \equiv \int \mathrm{d}^3r\, F(\bs{r}) \,\mathrm{e}^{-\mathrm{i}\bs{k} \cdot \bs{r}} \mathcal{L}_{\ell}(\hat{\bs{k}} \cdot \hat{\bs{r}})\,,
\end{equation}
where $F(\bs{r})$ is the weighted galaxy fluctuation field, i.e., the FKP field \citep{Feldman:1993ky}. Here, the line-of-sight is taken to align with the endpoint $\boldsymbol{\hat{r}}$ of the galaxy pair and $\mathcal{L}_{\ell}$ is the Legendre polynomial of order $\ell$. To be able to use Fast Fourier Transform (FFT) to speed up the calculation, $\hat{\bs{k}}$ and $\hat{\bs{r}}$ in $\mathcal{L}_{\ell}(\hat{\bs{k}} \cdot \hat{\bs{r}})$ must be decoupled. One can use either a Cartesian decomposition approach \citep{PhysRevD.92.083532, Bianchi_2015}, or a spherical harmonic decomposition method \citep{Hand_2017,2018MNRAS.473.2737S}. Since the spherical harmonics are a natural basis for decomposing functions on a sphere, the latter approach is much more efficient. Specifically,
\begin{equation}\label{eq:hand}
    \begin{aligned}
        F_{\ell}(\bs{k})
        &= \frac{4 \pi}{2 \ell+1} \sum_{m=-\ell}^{\ell} Y_{\ell}^{m*}(\hat{\bs{k}}) \int \mathrm{d}^3r \,F(\bs{r}) Y_{\ell}^{m}(\hat{\bs{r}}) \,\mathrm{e}^{-\mathrm{i} \bs{k} \cdot \bs{r}}\\
        &\equiv \frac{4 \pi}{2 \ell+1} \sum_{m=-\ell}^{\ell} f_{\ell}^{m}(\bs{k})\,,
    \end{aligned}
\end{equation}
where the second line defines $f_\ell^m(\bs{k})$. This means that the estimator presented in Eq.~(\ref{eq:Hand17}) requires only $(2\ell+1)$ FFTs to calculate the multipole of order $\ell$, compared to $(\ell+1)(\ell+2)/2$ FFTs needed with the Cartesian decomposition approach \citep{PhysRevD.92.083532,Bianchi_2015}.

The power spectrum multipoles are purely real quantities, although they are constructed from $2\ell+1$ spherical harmonic components. This reality condition, combined with the fact that $f_\ell^m(\bs{k})$ is the product of spherical harmonics $Y_\ell^{m*}(\hat{\bs{k}})$ and the spherical harmonic-weighted Fourier transform of a real-valued field, suggests that a further optimization is possible. Specifically, combining Eq.~(\ref{eq:hand}) and Eq.~(\ref{eq:key1}) yields the following symmetry relation:
\begin{equation}\label{eq:F_lm_SIC}
    f_{\ell}^{-m}(\bs{k})=(-1)^\ell f_{\ell}^{m*}(-\bs{k})\,.
\end{equation}
This implies that for each $F_\ell(\bs{k})$, only half of the $f_\ell^m(\bs{k})$ terms need to be computed directly---e.g., those with nonnegative $m$. The remaining terms can be derived via spatial inversion and complex conjugation, reducing the number of required FFTs $N_\mathrm{FFT}$ from $2\ell + 1$ to $\ell + 1$, yielding a factor of $\sim2$ speedup. For instance, obtaining the first three multipoles of $F_{\ell}(\bs{k})$ requires only $1+3+5=9$ FFTs, whereas the methods of \cite{Hand_2017} and \cite{PhysRevD.92.083532} require $15$ and $22$ FFTs, respectively.

However, modern power spectrum estimation codes such as \href{https://nbodykit.readthedocs.io/en/latest/}{\texttt{Nbodykit}} are MPI-based, with complex mesh grids distributed across multiple processors. These grids typically follow the standard memory layout: zero frequency first, followed by positive frequencies in increasing index, then negative frequencies wrapped at the end. Consequently, the spatial inversion operation $f_{\ell}^{m*}(-\bs{k})$ necessitates substantial data exchange among processors, making it nontrivial to implement efficiently. To circumvent this difficulty while preserving the theoretical speedup, we propose evaluating a modified quantity that avoids explicit spatial inversion. Instead of constructing $F_{\ell}(\bs{k})$ directly, we compute
\begin{equation}\label{eq:new_F_ell}
    \mathcal{G}_\ell(\bs{k})\equiv \frac{4\pi}{2\ell+1} \left[f_\ell^0(\bs{k})+2\sum_{m=1}^\ell f_\ell^m(\bs{k})\right]\,.
\end{equation}
When $\ell=0$, $\mathcal{G}_\ell(\bs{k})$ is identical to $F_0(\bs{k})$. Using Eq.~(\ref{eq:F_lm_SIC}), one can readily verify that
\begin{equation}\label{eq:F_and_new_G}
    F_\ell(\bs{k}) =\frac{1}{2}\left[\mathcal{G}_\ell(\bs{k}) + (-1)^\ell \mathcal{G}_\ell^*(-\bs{k})\right]\,.
\end{equation}
Substituting this identity into Eq.~(\ref{eq:Hand17}) yields
\begin{equation}
    \widehat{P}_{\ell}(k)=\frac{2 \ell+1}{I_{22}} \int \frac{\dd^2 \hat{k}}{4 \pi} \frac{F_0(\bs{k})}{2} \left[\mathcal{G}_\ell(\bs{k}) + (-1)^\ell \mathcal{G}_\ell^*(-\bs{k})\right]^*-N_{\ell}^{\mathrm{shot}}\,.
\end{equation}

Moreover, a change of variable from $\bs{k}$ to $-\bs{k}$ in the solid angle integration shows that
\begin{equation}\label{eq:mathcalF}
    \begin{aligned}
        \int \frac{\dd^2 \hat{k}}{4 \pi} F_0(\bs{k}) \left[\mathcal{G}_\ell^*(-\bs{k})\right]^*
        =
        \left[\int \frac{\dd^2 \hat{k}}{4 \pi} F_0(\bs{k}) \mathcal{G}^*_\ell(\bs{k})\right]^*\,.
    \end{aligned}
\end{equation}
Thus, the contribution from $\mathcal{G}_\ell^*(-\bs{k})$ to $\widehat{P}_{\ell}(k)$ is precisely the complex conjugate of the contribution from $\mathcal{G}_\ell(\bs{k})$. We therefore obtain the final estimator for even-order power spectrum multipoles:
\begin{equation}\label{eq:pk_final}
\boxed{
\widehat{P}_{\ell}^{\rm E}(k) =\mathbb{RE}\left[\frac{2 \ell+1}{I_{22}} \int \frac{\dd^2 \hat{k}}{4 \pi} F_0(\bs{k}) \mathcal{G}_{\ell}^{*}(\bs{k})\right] - N_\ell^\mathrm{shot}\,,
}
\end{equation}
where $\mathbb{RE}$ denotes taking the real part of the one-dimensional vector, and we have used the fact that $N_{\ell}^{\mathrm{shot}}\propto \delta_{\ell 0}^{\rm K}$. For odd $\ell$, replacing $\mathbb{RE}$ with $\mathbb{IM}$ (taking the imaginary part) yields the corresponding odd-multipole estimator. In this formulation, $N_{\rm FFT} = \ell+1$ is preserved without requiring any explicit spatial inversion operations.

\subsection{Scoccimarro estimator}\label{sec:l1_sco}

The Scoccimarro estimator \citep{PhysRevD.92.083532,2017MNRAS.465.1757G} projects the full bispectrum $\widehat{B}_{123}(k_1,k_2,k_3)$ onto Legendre polynomials $\mathcal{L}_\ell (\hat{q}_1 \cdot \hat{x}_1)$:
\begin{equation}\label{eq:bi_sco}
    \widehat{B}_{123}^{(\ell)}(k_1,k_2,k_3)= (2 \ell+1) \prod_{i=1}^3 \int_{k_i} {\rm d}^3 q_i\frac{\delta_{\mathrm{D}}(\bs{q}_{123})}{V_{123}^{\mathrm{T}} I_{33}}F_{\ell}(\bs{q}_1) F_0(\bs{q}_2) F_0(\bs{q}_3) -N_{123}^{(\ell)}\,.
\end{equation}
Here, the normalization factor is given by $I_{33} \equiv \alpha \sum_{j=1}^{N_r} \bar{n}^2(\bs{x}_j) w_j^3$ \citep{Scoccimarro_2000}, while the shot noise term $N_{123}^{(\ell)}$ and its acceleration method are presented in \S{\ref{sec:shot_sco}}. Since this estimator shares the $F_{\ell}(\bs{q})$ terms with the Yamamoto estimator, Eq.~(\ref{eq:F_and_new_G}) can again be used to reduce the number of FFTs. For the bispectrum, evaluating $F_{\ell}(\bs{q})$ via explicit spatial inversion (see Appendix~\ref{sec:space_inv}) instead of $\mathcal{G}_\ell(\bs{q}) $ offers several distinct advantages:

\begin{enumerate}
    \item \textbf{Negligible inversion cost.} The overhead of spatial inversion is dwarfed by the dominant costs of bispectrum measurement---namely, the numerous inverse FFTs and the triangle-closing operations.

    \item \textbf{Real-valued transforms and their benefits.} Because $F_{\ell}(\bs{q})$ is conjugate symmetric, its inverse Fourier transform $F_k^{(\ell)}(\bs{x}) \equiv \int_k {\rm d}^3 q \, \mathrm{e}^{{\mathrm i} \bs{q} \cdot \bs{x}} F_{\ell}(\bs{q})$ is strictly real (for even $\ell$) or purely imaginary (for odd $\ell$). In either case, we can store only the real or imaginary part and use c2r iFFTs (inverse real-to-complex FFTs), yielding three practical advantages:
\begin{itemize}
    \item \textbf{Memory saving:} halving the per-array memory footprint compared to storing complex $\mathcal{G}_\ell$-derived arrays.
    \item \textbf{Faster triangle closing:} the core integral $(2\pi)^{-3}\int {\rm d}^3 x \, F_{k_1}^{(\ell_1)}(\bs{x}) F_{k_2}^{(\ell_2)}(\bs{x}) F_{k_3}^{(\ell_3)}(\bs{x})$ involves only real multiplications, avoiding the higher cost of complex arithmetic.
    \item \textbf{Improved caching:} the reduced memory allows us to cache significantly more $k$-bin arrays, avoiding repeated iFFTs and thus leading to substantial speedups. For example, if the available memory enables caching twice as many iFFT kernels, the number of iFFTs that must be recomputed is reduced to roughly one quarter of the original cost for the three-variable Scoccimarro estimator, and to about one half for the two-variable Sugiyama estimator.
\end{itemize}
\end{enumerate}

\subsection{Sugiyama estimator}\label{sec:l1_sugi}

The Sugiyama estimator \citep{2019MNRAS.484..364S} decomposes the bispectrum into a tri-polar spherical harmonic (TripoSH) basis:
\begin{equation}\label{sugi_expand}
B\left(\bs{k}_1, \bs{k}_2, \hat{n}\right)=\sum_{\ell_1+\ell_2+L=\text{even}} B_{\ell_1 \ell_2 L}(k_1, k_2) S_{\ell_1 \ell_2 L}(\hat{k}_1, \hat{k}_2, \hat{n})\,,
\end{equation}
with the isotropic 3-point basis \citep{2022PNAS..11911366P,cahn2020isotropic}:
\begin{equation}\label{eq:sugi_base}
\begin{aligned}    
S_{\ell_1 \ell_2 L}(\hat{k}_1, \hat{k}_2, \hat{n}) 
= \frac{1}{H_{\ell_1 \ell_2 L}} \sum_{m_1 m_2 M}
\begin{pmatrix} \ell_1 & \ell_2 & L \\ m_1 & m_2 & M \end{pmatrix}  
y_{\ell_1}^{m_1}(\hat{k}_1) y_{\ell_2}^{m_2}(\hat{k}_2) y_L^M(\hat{n})\,,
\end{aligned}
\end{equation}
where $y_{\ell}^{m}=\sqrt{4\pi/(2\ell+1)}Y_{\ell}^{m}$, the $3\times 2$ matrix denotes the Wigner 3-$j$ symbol and $H_{\ell_1 \ell_2 L}=
\begin{pmatrix} \ell_1 & \ell_2 & L \\ 0 & 0 & 0 \end{pmatrix}$ enforces even $\ell_1+\ell_2+L$. Furthermore, under the plane-parallel approximation with axial symmetry about the line of sight, $L$ must be even, which implies $\ell_1+\ell_2$ is also even. Adopting the LOS as the direction to the third endpoint of the triangle, i.e., $\hat{x}\approx \hat{x}_3$, the estimator reads:

\begin{equation}\label{sugi_estimator}
\begin{aligned}
\widehat{B}_{\ell_{1} \ell_{2} L}(k_{1}, k_{2}) 
= \frac{N_{\ell_{1} \ell_{2} L}H_{\ell_{1} \ell_{2} L}}{I} \sum_{m_{1} m_{2} M}
\begin{pmatrix} \ell_{1} & \ell_{2} & L \\ m_{1} & m_{2} & M \end{pmatrix}  
\int {\rm d}^{3} x\, F_{\ell_{1}}^{m_{1}}(\bs{x}; k_{1}) F_{\ell_{2}}^{m_{2}}(\bs{x}; k_{2}) G_{L}^{M}(\bs{x})\,,
\end{aligned}
\end{equation}
where $I$ is the normalization factor and
\begin{equation}\label{F_and_G}
\begin{aligned}
F_{\ell}^m(\bs{x}; k) & = \int \frac{{\rm d}^2 \hat{k}}{4 \pi} \, \mathrm{e}^{\mathrm{i} \bs{k} \cdot \bs{x}} y_{\ell}^{m *}(\hat{k}) \frac{\delta n|_{\mathrm{FFT}}(\bs{k})}{W_{\text{mass}}(\bs{k})}\,, \\
G_L^M(\bs{x}) & = \int \frac{{\rm d}^3 k}{(2 \pi)^3} \, \mathrm{e}^{\mathrm{i} \bs{k} \cdot \bs{x}} \frac{\delta n_L^M|_{\mathrm{FFT}}(\bs{k})}{W_{\text{mass}}(\bs{k})}\,.
\end{aligned}
\end{equation}
The angular integrations represent an average over a spherical shell of midpoint radius $k$ and thickness $\Delta k$. In practice, this is discretized as a sum over Fourier modes satisfying $k - \Delta k/2 < |\bs{q}| < k + \Delta k/2$, normalized by the mode count $N_{\text{mode}}(k)$:
\begin{equation}
    \int \frac{{\rm d}^2 \hat{k}}{4 \pi}  = \frac{1}{N_{\text {mode }}(k)} \sum_{k-\Delta k / 2<|\bs{q}|<k+\Delta k/ 2}\,.
\end{equation}

To optimize this estimator, we exploit the symmetry of the Wigner 3-$j$ symbol~\citep{arfken2011mathematical}:
\begin{equation}\label{wigner_3j}
\begin{pmatrix} \ell_1 & \ell_2 & L \\ m_1 & m_2 & M \end{pmatrix}
= (-1)^{\ell_1+\ell_2+L}
\begin{pmatrix} \ell_1 & \ell_2 & L \\ -m_1 & -m_2 & -M \end{pmatrix}\,.
\end{equation}
Since Eq.~(\ref{eq:sugi_base}) restricts to $\ell_1+\ell_2+L = \text{even}$, the coupling coefficient is invariant under simultaneous sign flip of all magnetic quantum numbers. Thus sub-configurations naturally pair as $(m_1, m_2, M)$ and $(-m_1, -m_2, -M)$, with the sole exception of $\{0,0,0\}$. Writing the bispectrum as a sum over sub-configurations,
\begin{equation}
   \widehat{B}_{\ell_{1} \ell_{2} L}(k_{1}, k_{2}) 
   = \sum_{m_{1} m_{2} M} \widehat{B}_{\ell_{1} \ell_{2} L}^{m_1 m_2 M}(k_{1}, k_{2}),
\end{equation}
we show in Appendix~\ref{sec:proofs} that
\begin{equation}\label{eq:sugi_F_and_G}
\begin{aligned}
F_{\ell}^{-m}(\bs{x}; k) &= (-1)^{\ell+m} F_{\ell}^{m *}(\bs{x}; k)\,, \\ 
G_L^{-M}(\bs{x}) &= (-1)^M G_L^{M *}(\bs{x})\,.
\end{aligned}
\end{equation}
Combined with $m_1+m_2+M=0$ and $\ell_1+\ell_2 = \text{even}$, this yields
\begin{equation}\label{eq:sym_sugi}
\widehat{B}_{\ell_{1} \ell_{2} L}^{-m_1 -m_2 -M}(k_{1}, k_{2}) 
= \left[ \widehat{B}_{\ell_{1} \ell_{2} L}^{m_1 m_2 M}(k_{1}, k_{2}) \right]^*\,.
\end{equation}
Consequently, only half of the sub-configurations require explicit evaluation, reducing the computational cost by a factor of two. The optimized estimator takes the final form:
\begin{equation}
\boxed{\begin{aligned}
\widehat{B}_{\ell_{1} \ell_{2} L}(k_{1}, k_{2}) 
= \frac{H_{\ell_{1} \ell_{2} L} N_{\ell_{1} \ell_{2} L}}{I} \sum_{\mathcal{M}} 
\begin{pmatrix} \ell_1 & \ell_2 & L \\ m_1 & m_2 & M \end{pmatrix}
\times \eta \, \mathbb{RE} \left[ \int {\rm d}^{3} x\, F_{\ell_{1}}^{m_{1}}(\bs{x}; k_{1}) F_{\ell_{2}}^{m_{2}}(\bs{x}; k_{2}) G_{L}^{M}(\bs{x}) \right]\,,
\end{aligned}}
\end{equation}
where $\mathbb{RE}$ extracts the real part, $\mathcal{M}$ selects one element from each $\{(m_1,m_2,M),\,(-m_1,-m_2,-M)\}$ pair plus $\{0,0,0\}$,\footnote{The selection is not unique; one convenient choice is $M <0 $, plus $M = 0$ with $m_1 \le 0$.} and $\eta = 1$ for $\{0,0,0\}$ and $\eta = 2$ otherwise. This acceleration scheme for the Sugiyama estimator presented here has already been adopted in our collaborators' clustering measurement codes for DESI, such as \href{https://github.com/MikeSWang/Triumvirate/tree/main}{\texttt{Triumvirate}} and \href{https://github.com/adematti/jax-power}{\texttt{jax-power}}.

Further symmetries arise both between and within sub-configurations, but only for certain restricted angular-momentum combinations. 
For example, specific sub-configurations in multipoles of the form $\widehat{B}_{\ell \ell L}(k_1,k_2)$ benefit from an additional $\sim 2\times$ speedup through transpose and Hermitian symmetries. Combined with the generic $\sim 2\times$ acceleration discussed above, this results in an overall $\sim 4\times$ reduction in computational cost. All relevant symmetry relations are enumerated below.

\paragraph{Symmetry between sub-configurations}
\begin{equation}
    \widehat{B}_{\ell\ell L}^{m_1 m_2 M}(k_1,k_2)
    =
    \widehat{B}_{\ell\ell L}^{m_2 m_1 M}(k_2,k_1)\,.
\end{equation}

\paragraph{Symmetries within a single sub-configuration}
\begin{equation}
\begin{aligned}
    \widehat{B}_{\ell\ell L}^{m m M}(k_1,k_2)
    &= \widehat{B}_{\ell\ell L}^{m m M}(k_2,k_1)\,, \\
    \widehat{B}_{\ell\ell L}^{m -m 0}(k_1,k_2)
    &= \left[
    \widehat{B}_{\ell\ell L}^{m -m 0}(k_2,k_1)
    \right]^*\,.
\end{aligned}
\end{equation}

The shot-noise contribution to the bispectrum can be similarly optimized using the symmetry arguments developed here. Moreover, as we will discuss in \S{\ref{sec:sugi_shot}}, the distinctive structure of the shot noise admits even further simplifications.

\subsection{Comparison with real-form FFT acceleration}

An alternative acceleration strategy replaces the complex spherical harmonics $Y_\ell^m$ with their real-valued counterparts $X_{\ell m}$ (see Appendix~\ref{sec:ylm}) and employs real-to-complex (r2c) or complex-to-real (c2r) Fourier transforms. Since the galaxy density field is real, this also halves the computational redundancy.\footnote{In the wide-angle case, the \texttt{Nbodykit} implementation \citep{Hand_2017} has to default to complex FFTs, despite using real spherical harmonics, because its real-array mode assumes Hermitian symmetry for $F_{\ell}(\bs{k})$, whereas odd multipoles are anti-Hermitian. As a result, the factor-of-two reduction in computational cost is not realized in practice.}


For the Yamamoto and Scoccimarro estimators, this substitution is straightforward and amounts to a unitary transformation of the Legendre expansion. The Sugiyama estimator, however, presents complications. While multipoles with at least one zero index (e.g., $\widehat{B}_{202}$, $\widehat{B}_{220}$) reduce to Legendre expansions and remain compatible with real-form substitution, general multipoles (e.g., $\widehat{B}_{112}$, $\widehat{B}_{222}$) require constructing new isotropic basis vectors from tripolar real spherical harmonics. We discuss this in Appendix~\ref{sec:real_decomp}.

Despite the apparent symmetry of the two approaches---our method exploits symmetry \emph{across} sub-configurations, while rFFT exploits symmetry \emph{within} a single configuration---we do not recommend the rFFT approach for practical implementation, for two reasons:

\begin{enumerate}
    \item \textbf{Incomplete output handling.} The r2c transform produces only the non-redundant half of the Fourier-space grid. Binning operations subsequently require reconstructing the missing half, or more simply, the Hermitian symmetry weights in binning need to be treated with care, introducing additional bookkeeping operations.
    
    \item \textbf{Estimator fragmentation.} For the Sugiyama estimator, real-form substitution forces a case-by-case reformulation of the sub-configurations, breaking the uniformity of the original TripoSH decomposition. Furthermore, it would require maintaining a separate, parallel set of real-form utilities (e.g., binning and counting operations after r2c FFT, real spherical harmonic generators), fracturing the consistency of the codebase.
\end{enumerate}

Given these practical drawbacks, we adopt the complex-form optimization throughout this work.

\section{Modified high-order multipoles in power spectrum evaluation}\label{sec:level2}

The computational cost of the standard Yamamoto estimator increases significantly for higher-order multipoles as more FFTs are required to evaluate $F_\ell(\bs{k})$. However, this cost can be further reduced by exploiting the fact that higher-order Legendre polynomials are expressible algebraically as linear combinations of products of two lower-order Legendre polynomials. For example,
\begin{equation}\label{eq:legendre}
    \begin{aligned}
        \mathcal{L}_{4} &= \frac{35}{18}\mathcal{L}_{2}\cdot\mathcal{L}_{2}-\frac{5}{9}\mathcal{L}_{2}\cdot\mathcal{L}_{0}-\frac{7}{18}\mathcal{L}_{0}\cdot\mathcal{L}_{0}\,,\\
        \mathcal{L}_{6} &= \frac{11}{5}\mathcal{L}_{4}\cdot\mathcal{L}_{2}-\frac{4}{7}\mathcal{L}_{4}\cdot\mathcal{L}_{0}-\frac{22}{35}\mathcal{L}_{2}\cdot\mathcal{L}_{0}\,,\\
        \mathcal{L}_{8} &= \frac{1287}{490}\mathcal{L}_{4}\cdot\mathcal{L}_{4}-\frac{ 286}{245}\mathcal{L}_{4}\cdot\mathcal{L}_{2}-\frac{209}{1715}\mathcal{L}_{4}\cdot\mathcal{L}_{0}-\frac{78}{1715}\mathcal{L}_{2}\cdot \mathcal{L}_{0}-\frac{143}{490}\mathcal{L}_{0}\cdot\mathcal{L}_{0}\,.
    \end{aligned}
\end{equation}
If we assign two distinct line-of-sights to each Legendre polynomial pair, i.e., make the approximation $\mathcal{L}_{\ell_1}(\hat{\bs{k}}\cdot\hat{\bs{r}}_2)\cdot\mathcal{L}_{\ell_2}(\hat{\bs{k}}\cdot\hat{\bs{r}}_2) \approx \mathcal{L}_{\ell_1}(\hat{\bs{k}}\cdot\hat{\bs{r}}_1)\cdot\mathcal{L}_{\ell_2}(\hat{\bs{k}}\cdot\hat{\bs{r}}_2)$, lower-order $F_\ell(\bs{k})$ can be reused and power spectrum multipoles can therefore be estimated as \citep{PhysRevD.92.083532},
\begin{equation}\label{eq:Plnew}
    \begin{aligned}
        \widehat{P}_{4b} &= \frac{35}{18}\widehat{P}_{22} - \widehat{P}_2 - \frac{7}{2}\widehat{P}_0,\\
        \widehat{P}_{6b} &= \frac{11}{5}\widehat{P}_{42} - \frac{52}{63}\widehat{P}_4 - \frac{286}{175}\widehat{P}_2,\\
        \widehat{P}_{8b} &= \frac{1287}{490}\widehat{P}_{44} - \frac{374}{245}\widehat{P}_{42} - \frac{3553}{15435}\widehat{P}_4 - \frac{1326}{8575}\widehat{P}_2 - \frac{2431}{490}\widehat{P}_0\,.
    \end{aligned}
\end{equation}
Here we defined the two-indices power spectrum multipole estimator
\begin{equation}\label{eq:Pllp}
    \widehat{P}_{\ell \ell'} = \frac{2(\ell + \ell')+1}{I_{22}}\int\frac{\dd^2 \hat{k}}{4\pi}F_{\ell}(\bs{k})F_{\ell'}^*(\bs{k}) - N_{\ell \ell'}^{\rm shot}\, .
\end{equation}
 Note that this requires the traditional $F_\ell(\bs{k})$ rather than $\mathcal{G}_\ell(\bs{k})$; fortunately, $F_\ell(\bs{k})$ is readily obtained from $\mathcal{G}_\ell(\bs{k})$ using Eq.~(\ref{eq:F_and_new_G}). The shot-noise term is evaluated by
\begin{equation}
    N_{\ell \ell'}^{\rm{shot}} \equiv \frac{2(\ell + \ell')+1}{I_{22}}\left(\sum_{j=1}^{N_g}+\alpha^2 \sum_{j=1}^{N_r}\right) w_j^2 \int \frac{\dd^2 \hat{k}}{4 \pi} \mathcal{L}_{\ell}(\hat{\bs{k}} \cdot \hat{\bs{x}}_j)\mathcal{L}_{\ell'}(\hat{\bs{k}} \cdot \hat{\bs{x}}_j)\,,
\end{equation}
which yields $N_{22}^{\rm{shot}} = 9N_0/5$ and $N_{44}^{\rm{shot}} = 17N_0/9$.

This two-step procedure enables the evaluation of multipoles up to $\widehat{P}_{(2L) b}$ using only $\ell = L$ terms. Combined with the acceleration introduced in \S\ref{sec:level1}, we require only 4 FFTs for $\widehat{P}_{4b}$ and 9 FFTs for $\widehat{P}_{8b}$. 
In Table~\ref{tab:number_of_ffts}, the symbol $\ell_{\rm max}$ in the last column denotes the maximum multipole order we wish to evaluate. Our optimization for high-order multipoles applies exclusively to even $\ell$, while odd multipoles are not considered in this scheme.

It should be noticed that this idea itself is not new. This alternative estimator was proposed originally in \citet{PhysRevD.92.083532}, and has already been used in realistic galaxy surveys \citep[e.g.][]{Gil_Mar_n_2020}. The concerns regarding the use of this alternative estimator can be divided into two aspects. First, the author of \citet{PhysRevD.92.083532} points that the measured $\hat{P}_{4b}$ has larger variance compared to the traditional $\hat{P}_{4}$, indicating potential information loss. Second, as we use two distinct line-of-sights, this alternative estimator is expected to have slightly different wide-angle effects. In principle, this could be corrected by modifying the survey window matrix estimator following \citet{Beutler:2021eqq}, but this is beyond the scope of this paper, and we leave the complete treatment to future work. To address the above concerns, we perform validation tests on galaxy mocks in the next section.

\begin{table*}[!t]
    \centering
    \begin{tabular}{cccc}
    \hline
    \hline
    Estimator         & $\ell_{\rm max}=4$ & $\ell_{\rm max}=8$ & $\ell_{\rm max}=\ell$ \\
    \hline
    \cite{PhysRevD.92.083532} & 22 & 95 & $(\ell+2)(\ell+4)(2\ell+3)/24$ \\
    \cite{Hand_2017}          & 15 & 45 & $(\ell+1)(\ell+2)/2$ \\
    This work (\S{\ref{sec:level1}})       & 9  & 25 & $(\ell+2)^2/4$ \\
    This work (\S{\ref{sec:level1}}+\S{\ref{sec:level2}})     & 4  & 9  & $(\ell+4)^2/16$ \\
    \hline
    \end{tabular}
    \caption{\label{tab:number_of_ffts}Total number of FFTs required for measuring even power spectrum multipoles up to order $\ell_{\rm max}$ using different estimators.}
\end{table*}

\subsection{Validation}\label{sec:level2_validation}
We validate the approximation method using 2048 MultiDark-Patchy mock catalogs \citep{10.1093/mnras/stw1014, 10.1093/mnras/stv2826}, which reproduce both the survey geometry and the galaxy number density of the BOSS Data Release~12 CMASS sample \citep{pub.1038954845}. We adopt the North Galactic Cap (NGC) subsample in the redshift range $z \in [0.2,0.5]$, supplied with a random catalog 50 times denser than the galaxy catalog. The fiducial $\Lambda$CDM cosmology $(\Omega_{\mathrm{m}}=0.31)$ is used to convert equatorial coordinates to Cartesian coordinates. The catalog is interpolated onto a $512^3$ mesh of side length $3.5\gpcoh$ using a triangular-shaped cloud (TSC) assignment, yielding a Nyquist frequency of $k_{\rm N} = 0.46\hompc$. We use 30 uniform $k$-bins in the range $0 \le k \le 0.3\hompc$. Following \citet{Jing_2005} and \citet{2019MNRAS.484..364S}, we correct for the mass-assignment effect on the weighted overdensity field $\delta n_\ell^m(\boldsymbol{x}) = \delta n(\boldsymbol{x})Y_\ell^m(\hat{x})$ in Fourier space via
\begin{equation}
    \delta n_\ell^m(\bs{k}) = \frac{\mathcal{F}[\delta n_\ell^m(\bs{x})]}{W_{\text{mass}}^{\text{TSC}}(k)},
\end{equation}
where
\begin{equation}\label{compensation}
    W_{\text{mass}}^{\text{TSC}}(k) = \prod_{i=x, y, z} \left[\operatorname{sinc}\left(\frac{\pi k_i}{2 k_{\mathrm{N}}}\right)\right]^3.
\end{equation}

In addition to comparing the mean and standard deviation ratios between $\widehat{P}_{\ell b}$ and $\widehat{P}_{\ell}$, we also evaluate their specific signal-to-noise ratio (SNR), which serves, to some extent, as a proxy for the relative information content of the two estimators. Let $\boldsymbol{X}_i$ denote the power spectrum multipole vector measured from the $i$-th mock, and $\bar{\boldsymbol{X}}$ the mean vector averaged over all mocks. The covariance matrix is then
\begin{equation}
    \boldsymbol{C}= \frac{1}{N_{\rm mock}-1} \sum_{i=1}^{N_{\rm mock}}(\boldsymbol{X}_{i}-\bar{\boldsymbol{X}})(\boldsymbol{X}_{i}-\bar{\boldsymbol{X}})^{\rm T}\,,
\end{equation}
and the SNR is defined as
\begin{equation}
    \mathsf{SNR} = \sqrt{ f_{\rm H}\,\bar{\boldsymbol{X}}^{\rm T} \boldsymbol{C}^{-1}\bar{\boldsymbol{X}}}\,,
\end{equation}
where $f_{\rm H}$ is the Hartlap correction factor~\citep{Hartlap_2006},
\begin{equation}
    f_{\rm H} = \frac{N_{\rm mock}-p-2}{N_{\rm mock}-1}\,,
\end{equation}
with $N_{\rm mock}$ and $p$ denoting the number of mocks and the length of the data vector, respectively.

To further understand how differences in signal and statistical uncertainty between $\widehat{P}_{4b}$ and $\widehat{P}_{4}$ affect cosmological inference, we perform full-shape fits to the mean of 2048 Patchy mocks in the redshift-sliced NGC region, with covariance matrix rescaled by $1/15$. This is equivalent to increasing the effective survey volume by $15$, achieving $V_\mathrm{eff}\approx45\,[\mathrm{Gpc}/h]^3$ approximately DESI DR2 precision \citep{DESI:2025zgx}. We fit power spectrum monopole, quadrupole and hexadecapole within the $k$-range of $[0.02, 0.20]\,h\,\mathrm{Mpc}^{-1}$. Redshift-space power spectrum multipoles are modeled based on effective field theory (EFT) using the IR-resummed Eulerian flavor of Velocileptors code \citep{Chen:2020fxs,Maus:2024dzi}. Parameterizations and priors on EFT model parameters closely follow the DESI DR1 full-shape analysis \citep{DESI:2024jxi}, except that we do not fix high-order counter term parameter $\alpha_4$ and stochastic term parameter $\mathrm{SN}_4$ as we are including hexadecapole. We assume a flat $\Lambda$CDM cosmology with massless neutrinos, fixing power spectrum tilt $n_s$ to Patchy mocks' cosmology $n_s=0.9611$, along with a Gaussian BBN prior $\omega_b = 0.02214\pm0.00055$. Theoretical power spectrum multipoles are firstly evaluated for $\ell=0,2,4$ at the center of $k$-bins in the range $0<k<0.4\,h\,\mathrm{Mpc}^{-1}$ with bin width $\Delta k=0.001\,h\,\mathrm{Mpc}^{-1}$, and then are convolved with wide-angle matrix and survey window matrix calculated in \citet{Beutler:2021eqq} before being compared with the measured data. It should be noted that these wide-angle matrix and window matrix are derived for $\widehat{P}_4$. We expect $\widehat{P}_{4b}$ suffers from slightly different survey window and wide-angle effect, therefore in this sense the model we use for $\widehat{P}_{4b}$ is slight ``wrong''. However, the goal here is to study how this insufficient model affects the final cosmological result. 

\paragraph{Results} The mean and standard deviation of the traditional multipoles $P_\ell(k)$ are presented in Appendix~\ref{app:figures}. Figure~\ref{fig:P_ell_b} compares the mean values (left panel) and standard deviations (middle panel) of the modified estimators $\widehat{P}_{\ell b}$ with those of the traditional $\widehat{P}_\ell$. The mean values of $\widehat{P}_{4}$ and $\widehat{P}_{4 b}$ show good agreement for $k \ge 0.1 \, h\,\mathrm{Mpc}^{-1}$, while the modified estimators exhibit slightly larger variance. These results are nevertheless consistent with the findings of \citet{PhysRevD.92.083532}. Given that the redshift range adopted here is relatively low, the measurements are expected to be severely affected by the wide-angle effect; we therefore anticipate that this approximate estimator will perform better in higher-redshift samples. In the right panel, we compare the ratio of the SNR of the approximate estimator to that of the original estimator, i.e., $\mathsf{SNR}(\widehat{P}_{\ell b})/\mathsf{SNR}(\widehat{P}_{\ell})$. When only large-scale data (small $k$) are included, this ratio is noticeably below unity, suggesting a possible loss of information due to the approximation. As more small-scale data points are incorporated, the ratio rises steadily; notably, for the hexadecapole, it gradually approaches unity.

\begin{figure*}[!t]
\begin{center}
\includegraphics[width=0.5 \textwidth]{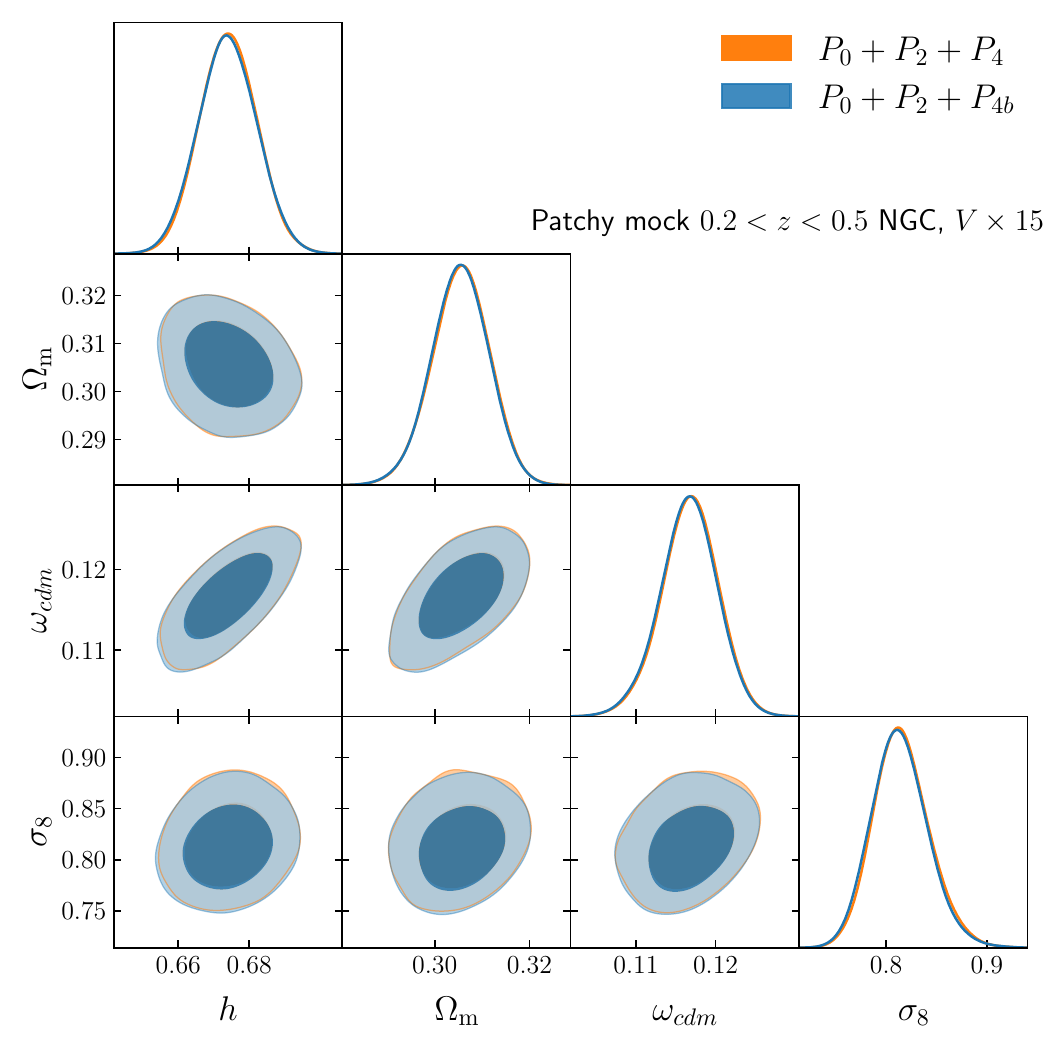}
\caption{\label{fig:p_ell_fitting}
Parameter fitting results using $\widehat{P}_0,\ \widehat{P}_2$, incorporating $\widehat{P}_4$ and its approximation $\widehat{P}_{4b}$.}
\end{center}
\end{figure*}

As shown in Fig.~\ref{fig:p_ell_fitting}, the posterior distributions of the cosmological parameters obtained from $\widehat{P}_{4b}$ and $\widehat{P}_{4}$ overlap almost perfectly, indicating that, at the precision level of DESI DR2, the difference between the two estimators has a negligible impact. Given the substantial computational speedup afforded by this approximation and the apparent minimal loss of information, we consider it a worthwhile trade-off.


\begin{figure*}[!t]
\begin{center}
\includegraphics[width=1.0 \textwidth]{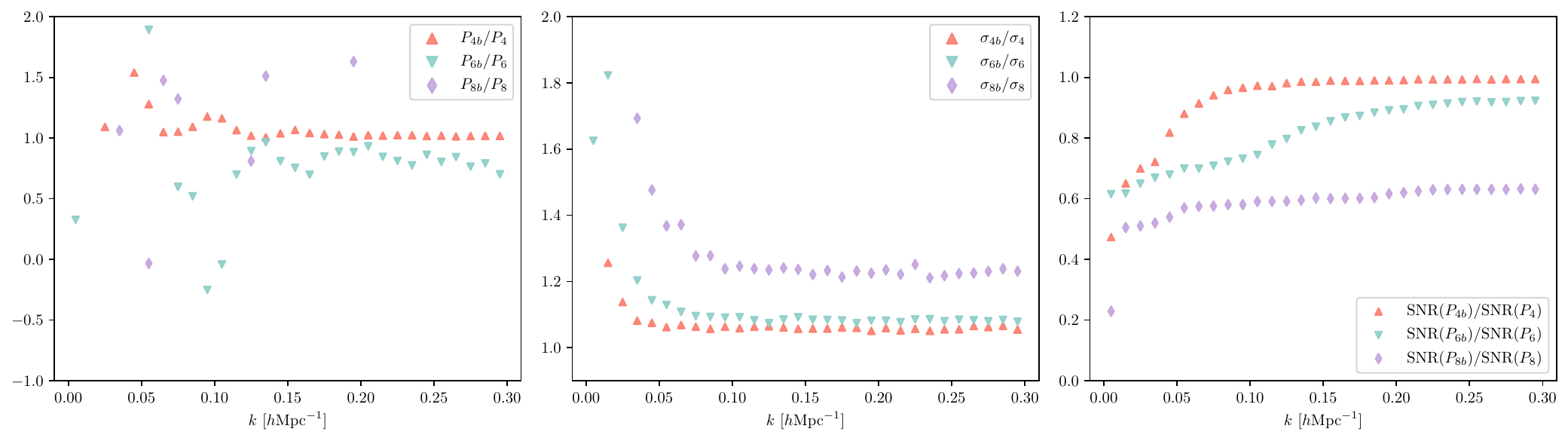}
\caption{\label{fig:P_ell_b}
From left to right: the mean ratio, standard deviation ratio, and SNR ratio of the shot-noise-subtracted power spectrum multipoles $\widehat{P}_{\ell b}$ to the standard ones $\widehat{P}_{\ell}$, computed from 2048 MultiDark-Patchy mocks of the NGC subsample in the redshift slice $z \in [0.2,0.5]$.}
\end{center}
\end{figure*}

\section{A New Compressed TripoSH Bispectrum Estimator}\label{sec:level3}

Comparing the Sugiyama estimator multipoles $\widehat{B}_{\ell 0 \ell}\left(k_{1}, k_{2}\right)$ with the Scoccimarro estimator $\widehat{B}_{\ell}\left(k_{1}, k_{2},k_{3}\right)$, one notices that the former can be viewed as the latter compressed along the $k_3$ direction, although each adopts a different default line of sight ($\hat{x}_3$ and $\hat{x}_1$, respectively). A natural generalization is to compress $\widehat{B}_{\ell}\left(k_{1}, k_{2},k_{3}\right)$ along another direction, e.g., $k_1$. In other words, the compression direction, the line of sight, and the $Y_{\ell}^m$ weights in $k$-space all coexist within a single field. To facilitate a direct comparison with the Sugiyama estimator and without loss of generality, we choose the compression direction to be $k_3$, yielding a new bispectrum estimator denoted as $\widehat{\mathcal{B}}$.
\begin{equation}\label{eq:new_bispec}
    \begin{aligned}
\widehat{\mathcal{B}}_{\ell_1 \ell_2 L}\left(k_1, k_2\right) & =\frac{H_{\ell_1 \ell_2 L} N_{\ell_1 \ell_2 L}}{I} \sum_{m_1 m_2 M}\begin{pmatrix}
 \ell_1 &  \ell_2 & L\\
  m_1 & m_2 & M
\end{pmatrix} 
 \int \frac{{\rm d}^2 \hat{k}_1}{4 \pi}  \int \frac{{\rm d}^2 \hat{k}_2}{4 \pi} \int {\rm d}^3 k_3 \\
& \times \delta_{\mathrm{D}}\left(\bs{k}_1+\bs{k}_2+\bs{k}_3\right)  y_{\ell_1}^{m_1*}(\hat{k}_3)y_{\ell_2}^{m_2 *}(\hat{k}_2) \delta n\left(\bs{k}_1\right) \delta n\left(\bs{k}_2\right) \delta n_L^M\left(\bs{k}_3\right)\,.
\end{aligned}
\end{equation}

Compared to the Sugiyama estimator, we have simply changed the argument of $y_{\ell_1}^{m_1*}$ from $\hat{k}_1$ to $\hat{k}_3$. This seemingly minor change leads to an important consequence: the monopole of the new estimator coincides with that of Sugiyama's, i.e., $\widehat{\mathcal{B}}_{000} = \widehat{{B}}_{000}$. We adopt this modification for two reasons. First, like the Sugiyama estimator, our estimator is expanded on an isotropic tripolar basis. Second, for special angular momentum configurations such as $\widehat{\mathcal{B}}_{\ell 0 \ell}$, only terms like $F_{0}^0 (\bs{x};k)$ are needed, avoiding a large number of iFFT calculations. Thus, while the new estimator can in principle be extended to arbitrary multipoles, we will \textbf{only} focus on comparing configurations of the form $\widehat{\mathcal{B}}_{\ell 0 \ell}$ and $\widehat{B}_{\ell 0 \ell}$ in what follows.

To better compare computational complexity, assume that in a single measurement, both $k_1$ and $k_2$ are divided into $N_{\rm b}$ bins. Ignoring for the moment shot noise and other FFT operations, whose cost scales as $2\ell+1$ (subdominant to the main cost). We consider only the FFTs required to obtain the binned kernels \(F_{\ell_1}^{m_1}(\bs{x};k_1)\) and \(F_{\ell_2}^{m_2}(\bs{x};k_2)\), which constitute the dominant computational cost. The number of FFTs\footnote{This is the minimum number of iFFTs required---specifically, $(2\ell+1)N_{\rm b}$ for $F_{\ell}^{m}(\bs{x}; k_{1})$ where $m = -\ell,-\ell+1,\dots,\ell$, and $N_{\rm b}$ for $F_{0}^{0}(\bs{x}; k_{2})$---based on the assumption that all these kernels can be cached and do not need to be recomputed. Once our lossless acceleration from \S{\ref{sec:level1}} is applied, this cost is reduced to $(\ell+2)N_{\rm b}$.} required for the Sugiyama \(\widehat{B}_{\ell0\ell}\) estimator scales as \((2\ell+2)N_{\rm b}\). In contrast, our $\widehat{\mathcal{B}}_{\ell 0 \ell}$ estimator requires only $N_{\rm b}$ FFTs. Since typically $N_{\rm b} \gg \ell$, the complexity of the remaining operations, which scale only with $\ell$ and are independent of $N_{\rm b}$, is negligible, our estimator is approximately $2\ell+2$ times faster than the Sugiyama estimator for $\widehat{{B}}_{\ell0\ell}$ measurements.

The shot noise contribution to this estimator is non-trivial; we present its mathematical form in \S{\ref{sec:new_shot}}.

\subsection{Validation}
\paragraph{Fisher matrix and mock setup} Despite the significant computational speedup offered by the new estimator, whether it retains as much information as the Sugiyama multipoles remains an open question. For brevity, we compare the quadrupole configurations of both estimators, i.e., $\widehat{{B}}_{202}$ and $\widehat{\mathcal{B}}_{202}$. We quantify the information content of different estimators by computing the Fisher information matrix,
\begin{equation}
F_{i j} = \frac{\partial \mu^{\rm T}}{\partial \theta_i} \bs{C}^{-1} \frac{\partial \mu}{\partial \theta_j}\,,
\end{equation}
where $\mu$ is the bispectrum data vector, $\theta = \{\Omega_m, \Omega_b, h, n_s, \sigma_8, M_\nu\}$ denotes the cosmological parameter vector, and $\bs{C}$ is the covariance matrix. Although it is indeed possible to compute the Fisher matrix analytically, this requires intricate theoretical modeling. A more straightforward approach is to use a set of mock data with varying cosmological parameters to compute the partial derivatives directly.

In this work, we use the \textbf{Molino} mocks~\citep{Hahn_2021}\footnote{\url{https://changhoonhahn.github.io/molino/current/}} to complete the validation. These mocks are constructed from the Quijote N-body simulations at $z=0$~\citep{Villaescusa_Navarro_2020} using the standard Halo Occupation Distribution (HOD) model. Specifically, we utilize 15,000 fiducial mocks with cosmology $\theta_{\rm fid} = \{0.3175, 0.049, 0.6711, 0.9624, 0.834, 0.0\}$ to estimate the covariance matrix, and an additional 14 suites of mocks, each containing 2,500 mocks, to derive the partial derivatives for the six cosmological parameters. For the first five parameters, the deviations from the fiducial values are $\{\pm 0.01, \pm 0.02, \pm 0.02, \pm 0.02, \pm 0.015\}$. The relevant partial derivatives are given by finite difference:
\begin{equation}
\frac{\partial \mu}{\partial \theta_i} \approx \frac{\mu(\theta_i^{+}) - \mu(\theta_i^{-})}{\theta_i^{+} - \theta_i^{-}}\,.
\end{equation}
In addition, four suites of mocks generated with the Zel'dovich approximation at $M_\nu = \{0.0, 0.1, 0.2, 0.4\}$ eV are used to obtain the partial derivative with respect to $M_\nu$~\citep{Hahn_2020}, specifically
\begin{equation}
\frac{\partial \mu}{\partial M_\nu} \approx \frac{-21\mu(\theta_{\rm fid}^{\rm ZA}) + 32\mu(M_\nu^{+}) - 12\mu(M_\nu^{++}) + \mu(M_\nu^{+++})}{1.2}\,.
\end{equation}

The galaxy catalogs in the Molino mocks have number density $n_g \sim 1.63\times 10^{-4}\,h^3{\rm Mpc}^{-3}$. Each catalog has volume $(1\,{\rm Gpc}/h)^3$ and contains $\sim 150,000$ galaxies.

We first add RSD to the galaxy catalogs with LOS $=(0,0,1)$, then use TSC assignment to interpolate them onto a $256^3$ mesh, yielding $k_{\rm N} = 0.804\,h{\rm Mpc}^{-1}$. Unlike the power spectrum, when measuring the bispectrum, we must ensure $k_{\rm max} < k_{\rm N}/2$. We therefore cut 15 bins uniformly in $k$-space over the range $0 \le k \le 0.3\,h{\rm Mpc}^{-1}$ to measure the two quadrupoles as well as the shared monopole. To reduce computational cost and the length of the data vector, we consider only the diagonal configuration ($k_1 = k_2$). We note that the subsequent Fisher information comparison is therefore limited to this diagonal subset.                

To avoid overestimating the information content of the Fisher matrix, we also apply the Hartlap correction to the inverse covariance matrix $\bs{C}^{-1}$, and then convert the Fisher matrices into cosmological parameter contours.

\begin{figure*}[!t]
\begin{center}
\includegraphics[width=0.8 \textwidth]{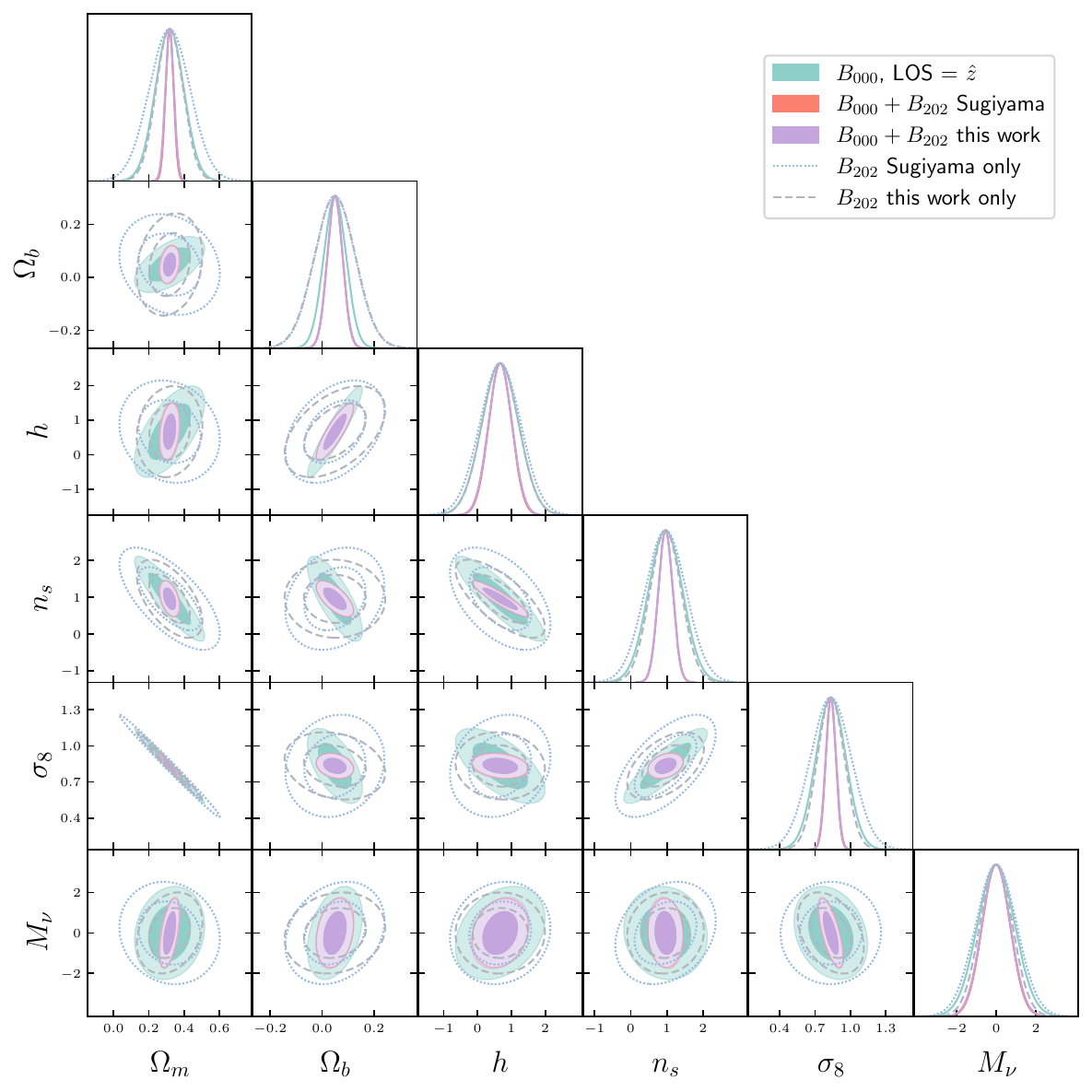}
\caption{\label{fig:fisher}
Cosmological parameter contours derived from the Fisher matrices of Sugiyama's $\widehat{B}_{202}$ and our $\widehat{\mathcal{B}}_{202}$. The blue dotted (no fill) and gray dashed (no fill) ellipses correspond to constraints from the two quadrupole estimators, respectively. The green-filled ellipse represents the constraints from their shared monopole. The orange- and purple-filled ellipses show the joint constraints obtained by combining the monopole with each of the two quadrupole estimators.}
\end{center}
\end{figure*}

\paragraph{Results} The bispectrum multipoles measured from the Molino mocks are shown in Appendix~\ref{app:figures} (Figure~\ref{fig:B_multipole}). In Figure~\ref{fig:fisher}, we present the two-dimensional parameter-ellipse constraints derived from the Fisher matrices of Sugiyama's $\widehat{B}_{202}$ and our $\widehat{\mathcal{B}}_{202}$. Our estimator performs at least as well as Sugiyama's for all parameters, and in most cases slightly better. However, when the shared monopole measurement is incorporated into the data vector, the joint constraining power of the two estimators tends to converge. There are two possible explanations for this. First, the long-axis direction of the ellipse for $\widehat{\mathcal{B}}_{202}$ aligns more closely with that of the monopole, suggesting a stronger information degeneracy between them. Second, although the quadrupole estimators differ, both encompass fluctuation information of the same order, so it is not surprising that their joint constraining results with the monopole are similar.

To summarise, the proposed $\widehat{\mathcal{B}}_{202}$ estimator achieves a level of information extraction on par with the Sugiyama quadrupole, yet delivers it at a markedly reduced computational expense. For upcoming galaxy surveys, this means that with a fixed set of mocks, our method requires far less computation. Alternatively, if the computational budget is held constant, it allows a substantially larger number of mocks to be processed—an increasingly important feature for robust covariance estimation.

\section{Bispectrum Shot Noise from Triangle-Constrained Spherical-Harmonic Integrals}\label{sec:shotnoise}

\subsection{Motivation} \label{sec:motivation_of_noise}
To achieve more precise measurements of the bispectrum, it is essential to carefully subtract the shot noise in the estimator. Unlike the constant shot noise present in the power spectrum, the shot noise of the bispectrum is no longer constant. Instead, it can be decomposed into three power-spectrum-like signal-noise cross terms and a constant pure Poisson noise term \citep{Scoccimarro_2000, Gil_Mar_n_2015}:
\begin{equation}
    B_{\text {shot }}\left (\bs{k}_1,\bs{k}_2,\bs{k}_3 \right ) =\frac{1}{\bar{n}}\left[P\left(\bs{k}_1\right)+P\left(\bs{k}_2\right)+P\left(\bs{k}_3\right)\right]+\frac{1}{\bar{n}^2}\,.
\end{equation}
where $\bar{n}$ is the mean number density of galaxies. However, when expanding into multipoles or compressing along a given $k$-axis, the practical shot noise evaluation can become non-trivial. Below we list several issues that need to be clarified, improved, or resolved in two existing bispectrum estimators as well as in our new estimator introduced in \S{\ref{sec:level3}}:
\begin{enumerate}
    \item \citet{PhysRevD.92.083532} gave an expression for the bispectrum shot noise, but the derivation was not fully presented, making it difficult to verify its correctness.
    \item \citet{2019MNRAS.484..364S} adopted the Rayleigh expansion of plane waves, which leads to a shot-noise expression that no longer retains the simple form analogous to power-spectrum estimators and introduces additional complexity.
    \item Our new bispectrum estimator introduced in \S{\ref{sec:level3}} requires a corresponding shot-noise subtraction, and the procedure cannot be directly borrowed from the approach of Sugiyama.
\end{enumerate}

Given these challenges, we aim for a more systematic and computationally efficient treatment. The core issues that arise during the calculation are similar across different estimators. We therefore focus on the following generic form and subsequently apply it to specific scenarios.

\subsection{A generic form of bispectrum shot noise under multipole expansion}
When computing the bispectrum expanded in a TripoSH basis as a sum of terms containing specific products of three spherical harmonics, e.g., $Y_{\ell_1}^{m_1}(\hat{q}_1)Y_{\ell_2}^{m_2}(\hat{q}_2)Y_{L}^{M}(\hat{x})$, the most critical step is to evaluate an integral of the form
\begin{equation}
\int \mathcal{D}\,Y_{\ell_1}^{m_1*}(\hat{q}_1)Y_{\ell_2}^{m_2*}(\hat{q}_2) \delta n(\bs{q}_1) \delta n (\bs{q}_2)\delta n_{L}^{M}(\bs{q}_3)\,,
\end{equation}
where $\delta n _{L}^{M}(\bs{q}) = \int {\rm d}^3 x \,{\mathrm e}^{-{\mathrm i}\bs{q}\cdot \bs{x}} \delta n(\bs{x})Y_{L}^{M*}(\hat{x})$ and $\int \mathcal{D} \equiv \prod_{i=1}^3 \int_{k_i}{\rm d}^3q_i \,\delta_{\rm D}(\bs{q}_1+\bs{q}_2+\bs{q}_3)$. A common strategy is to expand the Dirac delta function in the integral into plane waves,
$(2 \pi)^3 \delta_{\mathrm{D}}\left(\bs{q}_1+\bs{q}_2+\bs{q}_3\right)=\int {\rm d}^3 x \, \mathrm{e}^{\mathrm{i} x \cdot\left(\bs{q}_1+\bs{q}_2+\bs{q}_3\right)}$,
and then employ the Fast Fourier Transform to enforce the closure of $k$-space triangles. These FFTs constitute the main source of computational complexity when estimating bispectrum multipoles.

When calculating the shot noise, however, the reality of overlapping galaxies makes the situation quite different. Let the three galaxies in a given triplet be labeled $i,j,k$. The total shot noise receives contributions from four distinct cases: $i=j=k$, $i=j\neq k$, $i=k\neq j$, and $j=k\neq i$.
Following Eq.~(43) of \citet{2019MNRAS.484..364S} and omitting constant factors related to normalization and angular momenta, we can write the signal-noise cross term under the TripoSH expansion as
\begin{equation}\label{eq:generic_noise}
\begin{aligned}
        \mathcal{S} _{\ell_1\ell_2L} ^{m_1m_2M}|_{i=j\neq k}(k_1,k_2,k_3)
    &= \int \mathcal{D} \, Y_{\ell_1}^{m_1*}(\hat{q}_1)Y_{\ell_2}^{m_2*}(\hat{q}_2)
    \left(\sum_i^{N_g}+\alpha^2 \sum_i^{N_r}\right)\left[w\left(\boldsymbol{x}_i\right)\right]^2  \mathrm{e}^{-\mathrm{i} (\bs{q}_1+ \bs{q}_2)\cdot \boldsymbol{x}_i}
    \\
    &\quad\times \left(\sum_k^{N_g}-\alpha \sum_k^{N_r}\right)w\left(\boldsymbol{x}_k\right)Y_{L}^{M*}(\hat{x})  \mathrm{e}^{-\mathrm{i} \bs{q}_3\cdot \boldsymbol{x}_k}\\
    &\equiv \int \mathcal{D} \, Y_{\ell_1}^{m_1*}(\hat{q}_1)Y_{\ell_2}^{m_2*}(\hat{q}_2) N_{0}^{0*}(\bs{q}_3)\delta n_{L}^{M}(\bs{q}_3)\,,
\end{aligned}
\end{equation} 
where $\bs{x}_i$ and $w(\bs{x}_i)$ are the position and weight of a particular galaxy/random particle with index $i$ and 
\begin{equation}
\begin{aligned}
    \delta n_{\ell}^{m}(\bs{k}) &=  \left(\sum_i^{N_g}-\alpha \sum_i^{N_r}\right)w\left(\boldsymbol{x}_i\right)Y_{\ell}^{m*}(\hat{x}_i)  \mathrm{e}^{-\mathrm{i} \bs{k}\cdot \boldsymbol{x}_i}\,,\\
    N_\ell^m(\boldsymbol{k})&=\left(\sum_i^{N_g}+\alpha^2 \sum_i^{N_r}\right)\left[w\left(\boldsymbol{x}_i\right)\right]^2 Y_\ell^m\left(\hat{x}_i\right) \mathrm{e}^{-\mathrm{i} \boldsymbol{k} \cdot \boldsymbol{x}_i}\,.
\end{aligned}
\end{equation}
We have used the relation $\bs{q}_1+ \bs{q}_2 = -\bs{q}_3$, so that all information about the over-density we care about is absorbed into $N_{0}^{0*}(\bs{q}_3)\delta n_{L}^{M}(\bs{q}_3)$, while the integrand involving $\bs{q}_1$ and $\bs{q}_2$ consists solely of spherical harmonics.

A feasible approach would be to evaluate this expression in the same way as the signal part, using the plane-wave expansion of the Dirac function together with FFTs. However, this method has the drawback of doubling the number of required FFTs. \citet{2019MNRAS.484..364S} proposed an alternative solution, which replaces the FFT with a spherical-harmonic expansion of the plane waves (Eq.~\ref{eq:sph_wave_expan}), as shown in Eq.~(45) of \cite{2019MNRAS.484..364S}. Nevertheless, this approach still relies on spherical Bessel functions $j_{\ell}(kx)$, whose real-time evaluation remains relatively complex and offers only a limited reduction in computational cost. We therefore attempt to integrate analytically the part of the integral that contains only spherical harmonics.

\begin{figure}[!t]
\plotone{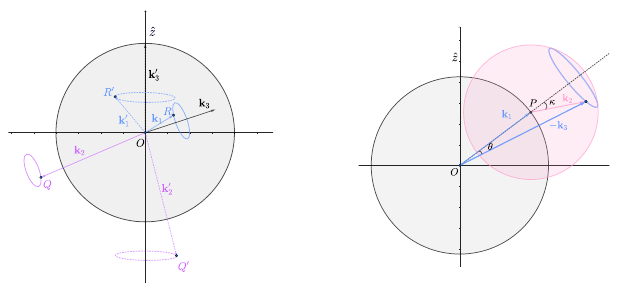}
\caption{Left: Schematic diagram of rigid body rotation of a triangle in Fourier space from its initial position to an arbitrary position. Right: Schematic diagram of spherical harmonic integration on an arbitrary two-dimensional spherical surface in three-dimensional space.}
\label{fig:sn_demo}
\end{figure}

We first rewrite the integral by eliminating the Dirac function from the integrand and instead expressing it as a multiple integral constrained by three specific integration domains. As illustrated in the left panel of Figure~\ref{fig:sn_demo}, once the three sides $\bs{k}_1,\bs{k}_2,\bs{k}_3$ of a triangle are fixed, the triangular constraint reduces the total number of degrees of freedom $V_{\rm T}$ below the direct product of the degrees of freedom of the three sides. For example, we can first choose one side, say $\bs{k}_3$, which is free to fill the entire spherical shell of radius $k_3$ and thickness $\delta k_3$. Its degree of freedom is therefore $V_{\rm sphere} = 4\pi k_3^2 \delta k_3$. Next, we choose the second side $\bs{k}_1$. Because the angle between $\bs{k}_1$ and $\bs{k}_3$ is fixed, $\bs{k}_1$ can only lie on a ring around $\bs{k}_3$ with thickness $\delta k_1$, yielding $V_{\rm ring} = 2\pi \sin{\theta_{13}}\,k_1 \delta k_1$. The third side $\bs{k}_2$ then has no angular freedom; it can only exist at a single point on any spherical surface. It does, however, retain some freedom in the radial direction. One can show that $V_{\rm dot} = \delta k_2 / \sin{\theta_{12}}$, and consequently
\begin{equation}
    V_{\rm T} \equiv V_{\rm sphere} V_{\rm ring} V_{\rm dot} = 8\pi^2 k_1 k_2 k_3 \,\delta k_1 \delta k_2 \delta k_3\,.
\end{equation}
In Appendix~\ref{sec:int_vol}, we provide a derivation of $V_{\rm dot}$ and $V_{\rm T}$.

With these considerations, we can rewrite Eq.~(\ref{eq:generic_noise}) as the
following thin-shell parametrization of the Dirac-delta constrained triangle
phase space:
\begin{equation}\label{eq:sn_form2}
    \mathcal{S} _{\ell_1\ell_2L} ^{m_1m_2M}|_{i=j\neq k}(k_1,k_2,k_3) =    \int_{V_{\rm ring}} {\rm d}^3 q_1 \,Y_{\ell_1}^{m_1*}(\hat{q}_1) \int_{V_{\rm dot}}{\rm d}^3 q_2 \,Y_{\ell_2}^{m_2*}(\hat{q}_2) \int_{V_{\rm sphere}}{\rm d}^3 q_3 \,N_{0}^{0*}(\bs{q}_3)\delta n_{L}^{M}(\bs{q}_3)\,.
\end{equation}
Here, the ring and dot domains should be understood as conditional domains fixed
by the same triangle geometry for each given \(\boldsymbol{q}_3\), rather than
as independent unconstrained integration domains. The other signal-noise cross terms, $ \mathcal{S} _{\ell_1\ell_2L} ^{m_1m_2M}|_{i=k\neq j}(k_1,k_2,k_3),\  \mathcal{S} _{\ell_1\ell_2L} ^{m_1m_2M}|_{j=k\neq i}(k_1,k_2,k_3)$ can be rewritten in a similar manner, while the Poisson noise term is simply a constant.


\subsection{Special integrals involving spherical harmonics}
\subsubsection{Lemma 1}
To evaluate Eq.~\eqref{eq:sn_form2}, we need to compute the following integral analytically:
\begin{equation}\label{eq:key_formula1}
    \mathcal{I}_{\ell_1\ell_2}^{m_1m_2}\equiv \frac{1}{V_{\rm dot}V_{\rm ring}}\int_{V_{\rm ring}} {\rm d}^3{k}_1 \,Y_{\ell_1}^{m_1}(\hat{k}_1)
\int_{V_{\rm dot}} {\rm d}^3{k}_2 \,Y_{\ell_2}^{m_2}(\hat{k}_2)\,.
\end{equation}

Let us adopt the thin-shell approximation and neglect the radial thickness, so that the integration region $V_{\rm ring}$ corresponds to the solid blue ring surrounding $\bs{k}_3$ in the left panel of Figure~\ref{fig:sn_demo}. This ring can be generated through the following rotational procedure: first, select an arbitrary triangular configuration $\left\{\bs{k}'_1,\bs{k}'_2,\bs{k}'_3\right\}$ as the initial state and designate the direction of $\bs{k}'_3$ as the $\hat{z}$-axis. Owing to the triangular constraint, $\bs{k}'_1$ can point to any point on the dashed blue ring in the left panel of Figure~\ref{fig:sn_demo}, denoted as $R'$. Correspondingly, $\bs{k}'_2$ will then point uniquely to $Q'$. One can readily see that $|\phi_{R'}-\phi_{Q'}| = \pi$.

Next, we use the Euler angles $(\alpha,\beta,\gamma)$ for rigid-body rotation to describe the rotation of the triangle.

\begin{enumerate}
    \item The triangle is rotated around the $\hat{z}$-axis by an azimuthal angle $\alpha$.
    \item The triangle is then rotated around the new $\hat{y}$-axis (resulting from the first rotation) by an angle $\beta$. At this stage, the vector $\bs{k}'_3$ is aligned with the target position $\bs{k}_3$, while the vectors $\bs{k}'_1$ and $\bs{k}'_2$ are rotated to points $R$ and $Q$, respectively, as illustrated in the diagram.
    \item Finally, rotating the triangle around the $\bs{k}_3$-axis through a full circle, parameterized by the angle $\gamma$, yields the target integration region.
\end{enumerate}

Note that since the positions of $\bs{k}_1$ and $\bs{k}_2$ correspond uniquely to each other, the angular degree of freedom of the integration region effectively resides on only one of the circular rings. We denote this sequence of rotations by $\hat{\mathcal{R}}(\alpha,\beta,\gamma)$, following the conventional Euler-angle parametrization with the $z$-$y$-$z$ rotation sequence, which must satisfy the conditions $\alpha,\gamma \in [0,2\pi]$, $\beta \in [0,\pi]$. In general, we can write this rotation process as
\begin{equation}
    \hat{\mathcal{R}}(\alpha,\beta,\gamma) \bs{k}'_i = \bs{k}_i\,, \quad i=1,2,3\,.
\end{equation}

In quantum mechanics, the Wigner $D$-functions relate the angular momentum eigenstates $\ket{\ell m}$ before and after a given rotation operation $\hat{\mathcal{R}}$ via
\begin{equation}\label{eq:rot_state}
\hat{\mathcal{R}}\ket{\ell m}
= \sum_n D_{nm}^{\ell}(\hat{\mathcal{R}})\ket{\ell n}\,.
\end{equation}
Furthermore, $Y_{\ell}^m(\hat{x})$ are the functional forms of $\ket{\ell m}$ in the spherical coordinate representation $\ket{\hat{x}}$.
We may therefore write the behavior of spherical harmonics under coordinate rotations described by the $D$-functions as
\begin{equation}\label{eq:rot_sph}
    Y_{\ell}^m(\hat{\mathcal{R}}\hat{x})=\sum_n D_{mn}^{\ell*}(\hat{\mathcal{R}})Y_{\ell}^n(\hat{x})\,.
\end{equation}
We refer the reader to Appendices~\ref{sec:wignerD} and~\ref{sec:rot_Ylm} for the definition and some commonly used properties of Wigner $D$-functions.
Inserting Eq.~\eqref{eq:rot_sph} into Eq.~\eqref{eq:key_formula1} and applying the thin-shell approximation once more, we obtain
\begin{equation}
\begin{aligned}
    \mathcal{I}_{\ell_1\ell_2}^{m_1m_2} 
&= \int_0^{2\pi} \frac{{\rm d}\gamma}{2\pi} \left[ \sum_{n_1=-\ell_1}^{\ell_1}Y_{\ell_1}^{n_1}(\theta_{R'},\phi_{R'})D^{\ell_1 *}_{m_1 n_1}(\alpha,\beta,\gamma)\right] \\
&\quad \times \left[ \sum_{n_2=-\ell_2}^{\ell_2}Y_{\ell_2}^{n_2}(\theta_{Q'},\phi_{Q'})D^{\ell_2 *}_{m_2 n_2}(\alpha,\beta,\gamma)\right]\,.
\end{aligned}
\end{equation}
Then, using the coupling rules for $D$-functions given in Appendix~\ref{sec:wignerD}, we find
\begin{equation}
\begin{aligned}
    \mathcal{I}_{\ell_1\ell_2}^{m_1m_2} 
&= \sum_{J=|\ell_1-\ell_2|}^{\ell_1+\ell_2} \sum_{n_1 n_2 M' N } (-1)^{M'+N}(2J+1)\left(\begin{array}{ccc}
\ell_1 & \ell_2 & J \\
m_1 & m_2 & -M'
\end{array}\right)\left(\begin{array}{ccc}
\ell_1 & \ell_2 & J  \\
n_1 & n_2 & -N
\end{array}\right) \\
&\quad \times\int_0^{2\pi} \frac{{\rm d}\gamma}{2\pi}  Y_{\ell_1}^{n_1}(\theta_{R'},\phi_{R'})Y_{\ell_2}^{n_2}(\theta_{Q'},\phi_{Q'}) {\rm{e}}^{{\mathrm i}M'\alpha }d_{M'N}^{J}(\beta){\rm{e}}^{{\mathrm i}N\gamma }\,.
\end{aligned}
\end{equation}
Note that $\int_{0}^{2\pi} {\rm{d}} \gamma \,{\rm{e}}^{{\mathrm i}N\gamma} = 2\pi \delta_{N0}^{\rm K}$, which immediately yields $N \equiv n_1+ n_2 = 0$. We therefore write $n = n_1$, with $n \in \left \{ -\ell ,\dots ,\ell-1,\ell \right \}$ and $\ell = \min\left \{ \ell_1,\ell_2 \right \}$. Furthermore,
${\rm{e}}^{{\mathrm i}M'\alpha }d_{M'0}^{J}(\beta) \equiv D_{M'0}^{J*}(\alpha,\beta,0)= \sqrt{4\pi/(2J+1)}\,Y_J^{M'}(\hat{k}_3)$. Recalling that $|\phi_{R'}-\phi_{Q'}| = \pi$, we finally obtain
\begin{equation}\label{eq:lemma1}
\boxed{
\begin{aligned}
\mathcal{I}_{\ell_1\ell_2}^{m_1m_2} (\hat{k}_3)
&=
\sum_{J=|\ell_1-\ell_2|}^{\ell_1+\ell_2} \sum_{M'n} (-1)^{M'+n} \sqrt{\frac{N_{\ell_1 \ell_2 J}}{4\pi}}\left(\begin{array}{ccc}
\ell_1 & \ell_2 & J \\
m_1 & m_2 & -M'
\end{array}\right)\left(\begin{array}{ccc}
\ell_1 & \ell_2 & J  \\
n & -n & 0
\end{array}\right)\\
&\quad \ \ \times \mathcal{L}_{\ell_1}^n(\cos \theta_{13}) \mathcal{L}_{\ell_2}^{-n}(\cos \theta_{23})  Y_J^{M'}(\hat{k}_3)\,,
\end{aligned}
}
\end{equation}
where $N_{\ell_1 \ell_2 J} \equiv (2\ell_1+1)(2\ell_2+1)(2J+1)$ and $\cos\theta_{ij} \equiv \hat{k}_i \cdot \hat{k}_j$. It is readily apparent that when the lengths of the three sides of a triangle are fixed, the associated Legendre function terms become constants. Consequently, the above integral depends solely on $\hat{k}_3$.

By considering special angular momentum configurations for $\mathcal{I}_{\ell_1\ell_2}^{m_1m_2}$, we can derive further interesting conclusions. First, setting $\ell_2=m_2 = 0$ leads to $n=0$, $J=\ell_1$, $m_1 = M'$, giving
\begin{equation}\label{eq:I_l0l}
    \mathcal I_{\ell0}^{m0} 
    =\frac{1}{\sqrt{4\pi} }{\int_{V_{\rm ring}} \frac{{\rm d}^3 k_1}{V_{\rm ring}} Y_{\ell}^{m}(\hat{k}_1)}
= \frac{1}{\sqrt{4\pi} }\mathcal{L}_{\ell}(\hat{k}_1\cdot \hat{k}_3 )\,Y_\ell^{m}(\hat{k}_3)\,.
\end{equation}
In this way, we have obtained an analytical expression for the integral of spherical harmonics over a specific ring in three-dimensional space, with the requirement that the ring's center--origin line is normal to its plane.

Next, we let $\ell_1 =\ell_2 =\ell$, $m_1 = -m_2 = m$, and $J=M'=0$, which yields
\begin{equation}\label{eq:I_ll0}
    \mathcal I_{\ell\ell}^{m-m}|_{J=0}^{M'=0}= \frac{(-1)^m}{4\pi}\mathcal{L}_\ell(\hat{k}_1\cdot \hat{k}_2)\,,
\end{equation}
where we have used Eq.~\eqref{eq:add_associated_L}, the addition theorem of associated Legendre functions. This shows that the integral is independent of both $m$ and $\hat{k}_3$, and implies that when integrating over the above region around the symmetry axis, additional implicit symmetries emerge in Eq.~\eqref{eq:I_ll0}.

\subsubsection{Lemma 2}
In the previous section, we proposed a new estimator $\widehat{\mathcal{B}}_{\ell0\ell}(k_1,k_2)$, which can be regarded as a compressed version of the Scoccimarro estimator. Consequently, the corresponding shot-noise calculation can also be written as the compressed form of Eq.~\eqref{eq:sn_form2}. As shown in the right panel of Figure~\ref{fig:sn_demo}, we now need to integrate $Y_{\ell}^m(-\bs{k}_3)$ over the sphere $S_P$ centered at $P$ with radius $k_2$. This sphere can be decomposed into a series of parallel rings sharing a common symmetry axis that passes through the origin. Using Eq.~\eqref{eq:I_l0l}, we obtain
\begin{equation}\label{eq:lemma2}
\boxed{\begin{aligned}
\int_{S_{P}} {\rm d}^2 S_p \,Y_\ell^{m}(-\hat{k}_3) 
=
4\pi k^2_2 \ g_{\ell}(t)\,Y_{\ell}^m(\hat{k}_1)\,,
 \end{aligned}}
\end{equation}
where $\int_{S_{P}}{\rm d}^2 S_p$ denotes integration over a sphere centered at $P$, while the integrand is a spherical harmonic whose argument is determined by the vector $-\bs{k}_3$\footnote{For convenience in the subsequent bispectrum calculations, we retain the variable as $-\bs{k}_3$. If one instead adopts the definition $\bs{k}_3 = \bs{k}_1 + \bs{k}_2$, the result takes the more compact form $\int_{S_{P}} {\rm d}^2 S_p \,Y_\ell^{m}(\hat{k}_3) 
=
4\pi k^2_2 \ g_{\ell}(t)\,Y_{\ell}^m(\hat{k}_1)$.}. 

Taking $t\equiv k_1/k_2$, the function $g_\ell(t)$ can be written as
\begin{equation}\label{eq:g_ell}
g_{\ell}(t)=
\begin{cases}
\displaystyle
\left[\int_0^{\kappa_0}+(-1)^\ell\int_{\kappa_0}^{\pi}\right]
\frac{{\rm d}\kappa}{2}\,\sin\kappa\,
\mathcal{L}_{\ell}\left(
\sqrt{\frac{\cos^2\kappa+t^2+2t\cos\kappa}
{1+t^2+2t\cos\kappa}}
\right)\,,
& 0\le t\le 1\,, \\[14pt]
\displaystyle
\int_0^\pi
\frac{{\rm d}\kappa}{2}\,\sin\kappa\,
\mathcal{L}_{\ell}\left(
\sqrt{\frac{\cos^2\kappa+t^2+2t\cos\kappa}
{1+t^2+2t\cos\kappa}}
\right)\,,
& t>1\,,
\end{cases}
\end{equation}
where $\kappa=\theta_{12}$ and $\theta=\pi-\theta_{13}$. For $0\le t\le 1$,
the transition point is $\kappa_0=\arccos(-t)$,
which corresponds to $\theta=\pi/2$. For $t>1$, no such transition point exists,
so the integral does not need to be split.

\begin{figure*}
\begin{center}
\includegraphics[width=0.6 \textwidth]{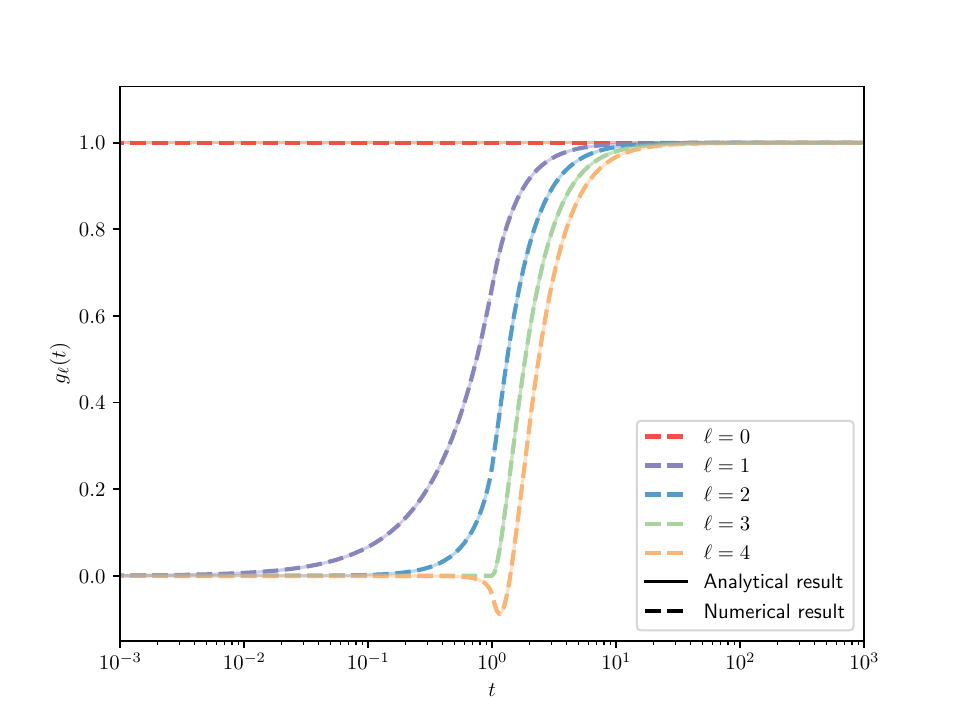}
\caption{\label{fig:g_ell}
First five orders of $g_{\ell}(t)$ given by Eq. \eqref{eq:g_ell} calculated numerically (dashed lines) and analytically (solid lines).}
\end{center}
\end{figure*}

We leave the derivation of Eq.~\eqref{eq:g_ell} and the analytical expressions for the first five orders of $g_{\ell}(t)$ to Appendix~\ref{sec:g_ell}. The function $g_\ell (t)$ can, of course, also be evaluated via numerical integration. In Figure~\ref{fig:g_ell}, we present both the analytical and numerical results for $g_\ell (t)$ for more general values of $t$, and find that they agree well with each other\footnote{When computing $g_{2}$ and $g_{4}$ using the analytical forms, both may exhibit divergence for $t < 10^{-3}$ due to insufficient default numerical precision in the program (we have verified that this occurs in both Python and Mathematica). In such cases, it is necessary to set the calculation precision manually.}. 
This result may be viewed as an off-centered generalization of the familiar
spherical average of spherical harmonics: the angular dependence is preserved
as \(Y_\ell^m(\hat{k}_1)\), while the displaced integration surface contributes
only through the scalar function \(g_\ell(t)\). The standard full-sphere
average in Eq.~\eqref{eq:single_ylm_integral} is then recovered in the
asymptotic limit \(t\to0\).

\subsection{Derivation of shot noise in three different bispectrum estimators}
\subsubsection{Scoccimarro estimator}\label{sec:shot_sco}
Using the special integrals derived in the previous section, we first calculate the relatively simpler shot noise of the Scoccimarro estimator. Following the notation of Eq. \eqref{eq:generic_noise}, we can readily derive $N^{(\ell)}_{i=j = k}$ and $N^{(\ell)}_{i\neq j = k}$, where $i,j,k$ label the three vertices in a galaxy triplet. These terms correspond to the first and third lines of Eq.~(58) in \cite{PhysRevD.92.083532}. However, the derivation of the second line in that equation is less straightforward. Based on the definition of the bispectrum shot noise, we present the following expression:
\begin{equation}
\begin{aligned}
        N^{(\ell)}_{i=k\neq j} 
        &= \frac{2\ell+1}{I_{33}V_{\mathrm{T}}}\sum_{m = -\ell}^{\ell}\prod_{n=1}^3 \int_{k_n} {\rm d}^3 q_n 
        F_0(\bs{q}_2)\left(\sum_{i=1}^{N_g}+\alpha^2 \sum_{i=1}^{N_r}\right) w_i^2 \, \mathrm{e}^{-{\mathrm{i}} (\bs{q}_1+\bs{q}_3) \cdot \bs{x}_i} 
        \mathcal{L}_\ell({\hat{q}_1 \cdot \hat{x}_i})\\
        & = \frac{4\pi}{I_{33}V_{\mathrm{T}}}\sum_{m = -\ell}^{\ell} \int_{V_{\rm ring}} {\rm d}^3{q}_1 Y_{\ell}^{m}(\bs{q}_1)
\int_{V_{\rm dot}} {\rm d}^3{q}_3 \int_{V_{\rm sphere}} {\rm d}^3{q}_2 F_0(\bs{q}_2) \left(\sum_{i=1}^{N_g}+\alpha^2 \sum_{i=1}^{N_r}\right) w_i^2 \, \mathrm{e}^{{\mathrm{i}} \bs{q}_2 \cdot \bs{x}_i} 
       Y_{\ell}^{m*}(\hat{x}_i)\,.
\end{aligned}
\end{equation}
Here $F_0(\bs{q}) = \int d^3 x \,\delta n(\bs{x})\mathrm{e}^{-{\mathrm i} \bs{q} \cdot \bs{x}}$. Inserting Eq.~\eqref{eq:I_l0l}, we obtain
\begin{equation}
\begin{aligned}
        N^{(\ell)}_{i=k\neq j} (k_2)
        &= \frac{2\ell+1}{I_{33}} \mathcal{L}_{\ell}(\hat{k}_1\cdot \hat{k}_2 )  \int_{k_2}  \frac{{\rm d}^3q_2}{N_{k_2}}  F_0(\bs{q}_2) F_{\ell}^{w*}(\bs{q}_2)\,,
\end{aligned}
\end{equation}
where $\int_{V_{\rm sphere}} {\rm d}^3q /V_{\mathrm{sphere}}= \int_{k} {\rm d}^3q /N_{k}$ and $F_{\ell}^w\left(\bs{q}\right) \equiv\left(\sum_{j=1}^{N_g}+\alpha^2 \sum_{j=1}^{N_r}\right) w_j^2 \, \mathrm{e}^{-{\mathrm i} \bs{q} \cdot \bs{x}_j} \mathcal{L}_{\ell}\left(\hat{q} \cdot \hat{x}_j\right)$,\footnote{Note that the mathematical form of the Fourier transform we use carries an extra negative sign in the exponent compared to that in Scoccimarro's paper.} which is precisely the shot noise given in the original paper. Similarly, we can derive $N^{(\ell)}_{i=j\neq k} (k_3)$. The lossless acceleration method we developed for power spectrum estimation can naturally be applied to the calculation of these shot noise terms.

\subsubsection{Sugiyama estimator}\label{sec:sugi_shot}
The derivations of the shot noise terms $S _{\ell_1\ell_2L}|_{i=j=k}$, $S _{\ell_1\ell_2L}|_{i\neq j=k}$, and $S _{\ell_1\ell_2L}|_{i=k\neq j}$ are straightforward. We therefore focus on applying the integral $\mathcal{I}_{\ell_1\ell_2}^{m_1m_2}$ derived in the previous section to the calculation of $S _{\ell_1\ell_2L}|_{i=j\neq k}(k_1,k_2)$. Adopting the notation of Eq. \eqref{eq:generic_noise}, we write,
\begin{equation}\label{eq:S3_sugi}
    \begin{aligned}
        S _{\ell_1\ell_2L}|_{i=j\neq k}(k_1,k_2) 
        &= \frac{H_{\ell_1\ell_2L}N _{\ell_1\ell_2L}}{I} \sum_{m_1 m_2 M}
        \begin{pmatrix}
        \ell_1  &\ell_2  & L\\
         m_1 & m_2 & M
         \end{pmatrix}
         \int\frac{{\rm d}^2{\hat{k}_1}}{4\pi} y_{\ell_1}^{m_1 *} (\hat{k}_1)\int\frac{{\rm d}^2{\hat{k}_2}}{4\pi} y_{\ell_2}^{m_2 *} (\hat{k}_2 )\\
         &\quad \times  \int {\rm d}^3 k_3 \,\delta_{\rm D}(\bs{k}_1+\bs{k}_2+\bs{k}_3)N_{0}^{0*}(\bs{k}_3)\delta n_{L}^{M}(\bs{k}_3)\,,
    \end{aligned}
\end{equation}
where again $y_{\ell}^{m}(\hat{k}) = \sqrt{{4\pi}/({2\ell+1})}Y_{\ell}^{m}(\hat{k})$ are the Racah-normalized spherical harmonics; $\delta_{\ell}^{m}(\bs{k}),\ N_{\ell}^{m}(\bs{k})$ are also defined using this normalization. Evaluating the integrals over $\bs{k}_1$ and $\bs{k}_2$ in the above equation using FFTs is straightforward: one first expands the Dirac delta into plane waves and then employs the identity
\begin{equation}
    \int \frac{{\rm d}^2 \hat{k}}{4 \pi} y_{\ell}^{m *}(\hat{k})\mathrm{e}^{{\mathrm i} \bs{k} \cdot \bs{x}} = \frac{V_{\rm mesh}}{N_{\rm mode}} \int\frac{{\rm d}^3k}{(2\pi)^3}\Phi(k)y_{\ell}^{m *}(\hat{k})\mathrm{e}^{{\mathrm i} \bs{k} \cdot \bs{x}}\,,
\end{equation}
where $V_{\rm mesh}$ is the volume of the interpolated mesh and $\Phi(k)$ is the radial binning mask defined in Eq.~\eqref{eq:binning_mask}. Since this FFT-based method for computing the shot noise follows a procedure that is nearly identical to the signal calculation and does not rely on any radial thin-shell approximation, we can treat it as a reference for assessing the reliability of other approximate methods in the subsequent analysis, despite its considerable computational cost.

To avoid performing these expensive FFTs, \cite{2019MNRAS.484..364S} employed the relation (see Eq.~\ref{eq:sugi_ylm})
\begin{equation}
    \int \frac{{\rm d}^2 \hat{k}}{4 \pi} y_{\ell}^{m *}(\hat{k})\mathrm{e}^{{\mathrm i} \bs{k} \cdot \bs{x}} = {\mathrm i}^{\ell} j_{\ell}(k x)y_{{\ell} m}^*(\hat{{x}})\,.
\end{equation}
This approach replaces the complex FFT operations with calculations involving spherical Bessel functions and spherical harmonics. However, due to the non-negligible thickness of each bin shell, the parameter $k$ in $j_{\ell}(kx)$ can only be interpreted as a characteristic value associated with that shell---such as its central value or the mean radial distance of the bin. Hence, although this method is relatively inexpensive compared to the FFT approach, it remains only a thin-shell approximation.

In order to perform this part of the calculation more efficiently, we first rewrite Eq.~\eqref{eq:S3_sugi} as integrals over several specific integration regions:
\begin{equation}\label{eq:S3_sugi2}
    \begin{aligned}
S _{\ell_1\ell_2L}|_{i=j\neq k}(k_1,k_2) 
        &= \frac{H_{\ell_1\ell_2L}N _{\ell_1\ell_2L}}{IV_{\rm C}} \sum_{m_1 m_2 M}\left(\begin{array}{ccc}
\ell_1 & \ell_2 & L \\
m_1 & m_2 & M
\end{array}\right) 
\int_{|k_1-k_2|}^{k_1+k_2} {\rm d}k_3 \int {\rm d}^2\hat{k}_3
\,\delta n_{L}^M(\bs{k}_3) N_0^{0*}(\bs{k}_3)\\
& \quad\times
\int_{V_{\rm ring}} {\rm d}^3{k}_1 y_{\ell_1}^{m_1*}(\hat{k}_1)
\int_{V_{\rm dot}} {\rm d}^3{k}_2 y_{\ell_2}^{m_2*}(\hat{k}_2)\,,
    \end{aligned}
\end{equation}
where $V_{\rm C}$ is the total effective integration volume (see Appendix~\ref{sec:int_vol}). Note that in this estimator, the bin center $k_3$ can vary over different discrete values; therefore, for any given $k_3$, the sizes of $V_{\rm ring}({k}_1)$ and $V_{\rm dot}({k}_2)$ will change accordingly. One can verify that $V_{\rm ring}({k}_1) V_{\rm dot}({k}_2) \propto 1/k_3$.

Substituting Eq.~\eqref{eq:lemma1} into Eq.~\eqref{eq:S3_sugi2} and utilizing the orthonormality of the 3-$j$ symbols given by Eq.~\eqref{eq:3j-orthonormal}, we obtain
\begin{equation}\label{eq:S3_sugi_new}
\boxed{
\begin{aligned}
S _{\ell_1\ell_2L}|_{i=j\neq k}(k_1,k_2)
= \frac{H_{\ell_1\ell_2L}N _{\ell_1\ell_2L}}{I}
\frac{\sum_{k_{3} =k_{\rm 3, min}}^{k_{\rm 3,max}}q_{\ell_1\ell_2L}(k_1,k_2,k_{3}) \mathcal{Q}_{L}(k_{3})/k_{3}}{\sum_{k_{3}=k_{\rm 3,min}}^{k_{\rm 3,max}}N_{\rm mode}(k_{3})/k_{3}}\,,
\end{aligned}
}
\end{equation}
with 
\begin{equation}\label{eq:q_ells}
\begin{aligned}
q_{\ell_1\ell_2L}(k_1,k_2,k_{3})=\sum_n(-1)^n\mathcal{L}_{\ell_1}^n(\hat{k}_1 \cdot\hat{k}_{3}) \mathcal{L}_{\ell_2}^{-n}(\hat{k}_2 \cdot\hat{k}_{3})
\left(\begin{array}{ccc}
\ell_1 & \ell_2 & L  \\
n & -n & 0 
\end{array}\right)\,,
\end{aligned}
\end{equation}
and
\begin{equation}\label{eq:Q_ell}
\begin{aligned}
\mathcal{Q}_{\ell}(k)   =  \frac{1}{N_{\rm mode}(k)} \sum_{k-\Delta k / 2< |\bs{q}| <k+\Delta k / 2} \sum_{m=-\ell}^{\ell} y_{\ell}^{m}(\hat{q})
    \left[\delta n_{\ell}^m(\bs{q}) N_0^{0*}(\bs{q})-\frac{\bar{S}_\ell^m C_{\rm shot}(\bs{q})}{W^2_{\rm mass}(\bs{q})}\right]\,.
\end{aligned}
\end{equation}
Here the minimum and maximum values that $k_3$ can take, denoted $k_{3,\rm min}$ and $k_{3,\rm max}$ respectively, are given by $|k_1 - k_2| + \Delta k / 2$ and $k_1 + k_2 - \Delta k / 2$, where $k_1$ and $k_2$ are fixed bin centers. In the calculation of $\mathcal{Q}_{\ell}(k)$ we include the shot-noise correction term ${\bar{S}_\ell^m C_{\rm shot}(\bs{q})}/{W^2_{\rm mass}(\bs{q})}$. The definitions of $\bar{S}_L^M$, $C_{\rm shot}(\bs{q})$, and $W_{\rm mass}(\bs{q})$ can be found in \cite{2019MNRAS.484..364S}.

For completeness, we also present optimized expressions for the remaining three shot noise terms:
\begin{equation}\label{sugi_shot}
    \boxed{
    \begin{aligned}
&\left.S_{\ell_1 \ell_2 L}\right|_{i=j=k}  =\delta_{\ell_1 0}^{\mathrm{K}} \delta_{\ell_2 0}^{\mathrm{K}} \delta_{L 0}^{\mathrm{K}}(1 / I) \bar{S}_{L=0}^{M=0}\,, \\
&\left.S_{\ell_1 \ell_2 L}\right|_{i \neq j=k}\left(k_1\right)  =\delta_{\ell_1 L}^{\mathrm{K}} \delta_{\ell_2 0}^{\mathrm{K}} \frac{2 L+1}{I}\mathcal{H}_{L}(k_1)\,,\\
&\left.S_{\ell_1 \ell_2 L}\right|_{i = k\neq j}\left(k_2\right)  =\delta_{\ell_2 L}^{\mathrm{K}} \delta_{\ell_1 0}^{\mathrm{K}} \frac{2 L+1}{I}\mathcal{H}_{L}(k_2)\,,
\end{aligned}
    }
\end{equation}
where we define
\begin{equation}\label{eq:H_ell}
\begin{aligned}
\mathcal{H}_{\ell}(k)   =   \frac{1}{N_{\rm mode}(k)} \sum_{k-\Delta k / 2< |\bs{q}| <k+\Delta k / 2} \sum_{m=-\ell}^{\ell}  y_{\ell}^{m}(\hat{q})
    \left[\delta n(\bs{q}) N_\ell^{m*}(\bs{q})-\frac{\bar{S}_\ell^m C_{\rm shot}(\bs{q})}{W^2_{\rm mass}(\bs{q})}\right]\,.
\end{aligned}
\end{equation}
Again, the acceleration method developed in \S{\ref{sec:level1}} can be applied to speed up the computation of both $\mathcal{H}_{\ell}$ and $\mathcal{Q}_{\ell}$.

In the next section, we will verify the accuracy of this method using analytical integration and evaluate the speed of several computational approaches.

\subsubsection{Our efficient new estimator}\label{sec:new_shot}
To subtract the corresponding shot noise $\mathcal{N}_{\ell 0 \ell}$ from the signal of our new estimator $\widehat{\mathcal{B}}_{\ell 0 \ell}$, we follow a procedure similar to the previous one and write
\begin{equation}\label{eq:S1_new_intermediate}
    \begin{aligned}
\mathcal{N}_{\ell 0 \ell}|_{i\neq j = k}(k_1,k_2) 
&= \frac{2\ell+1}{I}\sum_{m=-\ell}^{\ell} \int\frac{{\rm d}^2{\hat{k}_1}}{4\pi} \int\frac{{\rm d}^2{\hat{k}_2}}{4\pi}
\int {\rm d}^3 k_3 \,\delta_{\rm D}(\bs{k}_1+\bs{k}_2+\bs{k}_3)
\delta n(\bs{k}_1) N_\ell^{m*}(\bs{k}_1) y_{\ell}^{m } (\hat{k}_3)\\
        &= (-1)^{\ell}\frac{2\ell+1}{I}\sum_{m=-\ell}^{\ell}\int  \frac{{\rm d}^3 k_1}{N_{k_1}} \delta n(\bs{k}_1) N_\ell^{m*}(\bs{k}_1) \int_{S_P} \frac{{\rm d}^3 k_3}{N_{k_2}} y_\ell^{m}(-\hat{k}_3)\,.
 \end{aligned}
\end{equation}
Substituting Eq.~\eqref{eq:lemma2} into Eq.~\eqref{eq:S1_new_intermediate}, we obtain
\begin{equation}\label{eq:S1_new}
\begin{aligned}
\mathcal{N}_{\ell 0 \ell}|_{i\neq j = k}(k_1,k_2) 
=(-1)^\ell g_{\ell}(t) \frac{2 \ell+1}{I}\int \frac{{\rm d}^3 {k}_1}{N_{k_1}} \sum_{m=-\ell}^\ell  y_\ell^m(\hat{k}_1) \delta n(\bs{k}_1) N_\ell^{m*}(\bs{k}_1)\,.
 \end{aligned}
\end{equation}
Similarly, we can deduce $\mathcal{N}_{\ell 0 \ell}|_{i=k\neq j}(k_1,k_2)$. The term $\mathcal{N}_{\ell 0 \ell}|_{i=j\neq k}(k_1,k_2)$ can be derived in a manner analogous to that used for $S_{\ell 0 \ell}|_{i=j\neq k}(k_1,k_2)$. Finally, we present the full set:
\begin{equation}
    \boxed{
    \begin{aligned}
      &\mathcal{N}_{\ell 0 \ell}|_{i= j = k} =  \delta_{\ell 0}^{\rm K}\bar{S}_{0}^{0}\,,\\
      &\mathcal{N}_{\ell 0 \ell}|_{i\neq j = k}(k_1,k_2) = (-1)^\ell g_{\ell}(t)\frac{2 \ell+1}{I}\mathcal{H}_{\ell}(k_1)\,,\\
     & \mathcal{N}_{\ell 0 \ell}|_{i=k\neq j}(k_1,k_2)=
     (-1)^\ell g_{\ell}(1/t)\frac{2 \ell+1}{I}\mathcal{H}_{\ell}(k_2)\,,\\
     &\mathcal{N} _{\ell0\ell}|_{i=j\neq k}(k_1,k_2)  = 
  \frac{2\ell+1}{I}
\frac{\sum_{k_{3} =k_{\rm 3,min}}^{k_{\rm 3,max}}\mathcal{Q}_{\ell}(k_{3})/k_{3} }{\sum_{k_{3}=k_{\rm 3,min}}^{k_{\rm 3,max}}N_{\rm mode}(k_{3})/k_{3}}\,.
    \end{aligned}
    }
\end{equation}
One can quickly see that ${\mathcal{N}}_{\ell0\ell} |_{i\neq j=k}(k_1) =(-1)^{\ell}g_{\ell}(t)S_{\ell 0 \ell}|_{i\neq j=k}(k_1)$. According to Eq.~\eqref{sugi_shot}, $\left.S_{\ell_1 \ell_2 L}\right|_{i = k\neq j}\left(k_1\right)$ vanishes when $\ell \neq 0$, whereas ${\mathcal{N}}_{\ell0\ell} |_{i=k\neq j}(k_2)$ remains non-zero.

\subsection{Validation of the analytical method in the Sugiyama estimator}

\subsubsection{Methodology}

We compare three methods for computing the bispectrum shot noise component $S_{\ell_1 \ell_2 L}|_{i=j\neq k}(k_1,k_2)$: the analytical method based on Eq.~(\ref{eq:S3_sugi_new}), the FFT method, and the original Sugiyama method employing the plane-wave Rayleigh expansion. To evaluate their accuracy and precision, we measure the diagonal bispectrum multipoles $\widehat{B}_{\ell_1 \ell_2 L}(k,k)$, along with $S_{\ell_1 \ell_2 L}|_{i=j\neq k}(k,k)$ computed by all three methods in 3 most commonly used low-order cases: $\ell_1 \ell_2 L = \{000, 202, 220\}$, using the 200 MultiDark-Patchy mocks from Section~\ref{sec:level2}. We also select the CMASS North Galactic Cap (NGC) samples over the redshift range $0.43 < z < 0.75$. The box size, assignment function, and compensation scheme remain identical to those in Section~\ref{sec:level2}. The $k$-space is uniformly divided into 20 bins over $0 \le k \le 0.2\ h\text{Mpc}^{-1}$. To investigate the impact of bin width on the two approximate methods, we additionally measure $S_{220}|_{i=j\neq k}(k,k)$ using 10 and 40 bins for all three methods, though without computing the full bispectrum multipoles.

For computational speed evaluation, we conduct dedicated timing measurements with multi-threading acceleration disabled for all FFT operations, ensuring that the advantage of our new method in reducing the number of required FFTs is fairly assessed. All three methods are run simultaneously on identical hardware and software platforms to avoid performance variations due to differences in supercomputing cores.

\subsubsection{Results}

\begin{figure*}
\begin{center}
\includegraphics[width=1 \textwidth]{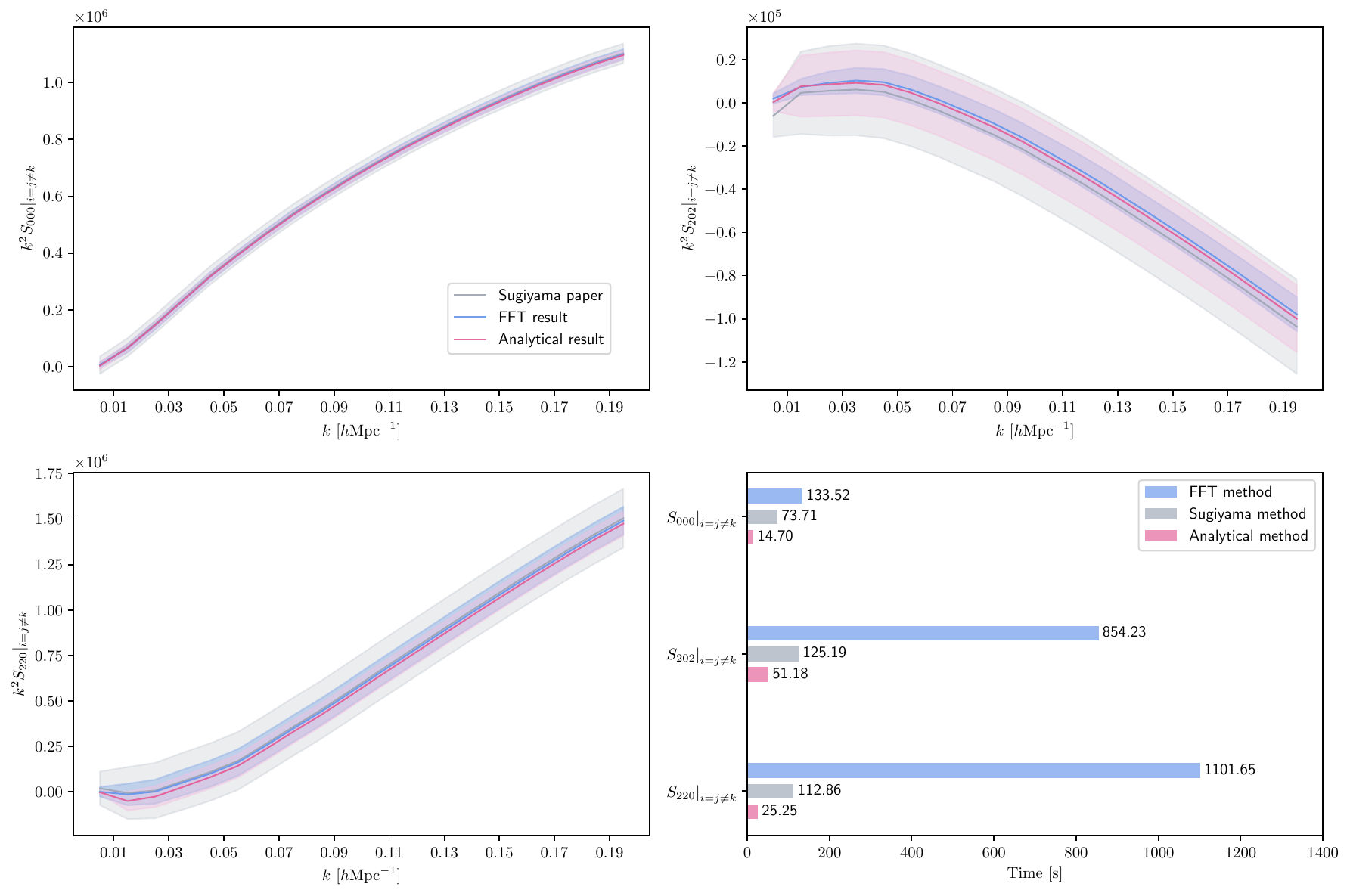}
\caption{\label{fig:shot_S3}
Top left: Shot noise term $S_{000}|_{i=j\neq k}$ scaled by $k^2$, evaluated from MD-patchy mocks using Sugiyama's original method, the FFT method, and the analytical method, respectively. The solid lines represent the mean value, while the colored bands indicate the standard deviation across 200 mocks. Top right and bottom left: Same format as top left, but for $S_{202}|_{i=j\neq k}$ and $S_{220}|_{i=j\neq k}$, respectively. Bottom right: Speed performance of the three methods, benchmarked on the same hardware platform with comparable code implementations.
}
\end{center}
\end{figure*}

We begin by examining the performance of the three methods in terms of accuracy, precision, and computational speed, as summarized in Figure~\ref{fig:shot_S3}. The first three subplots show the mean values (solid lines) and standard deviations (shaded areas) of $S_{000}|_{i=j\neq k}(k,k)$, $S_{202}|_{i=j\neq k}(k,k)$, and $S_{220}|_{i=j\neq k}(k,k)$ measured from 200 mocks. The bottom-right subplot presents the runtime comparison under identical hardware and software conditions, with multi-threading disabled to fairly assess the intrinsic computational cost. Several key observations emerge. First, the analytical method is substantially faster than both the FFT and Sugiyama methods, and its speed advantage becomes even more pronounced when computing the full 2D bispectrum multipoles. Second, for all three multipole configurations, the standard deviation of the analytical method closely matches that of the FFT method (the reference standard) and is considerably smaller than that of the Sugiyama method. Third, in terms of accuracy, the analytical method yields mean values that are nearly unbiased for $B_{000}$ and $B_{202}$, closely tracking the FFT reference. However, for $B_{220}$, the analytical method exhibits a noticeable offset relative to the FFT method, suggesting a configuration-dependent bias that warrants further investigation.

\begin{figure*}
\begin{center}
\includegraphics[width=0.9 \textwidth]{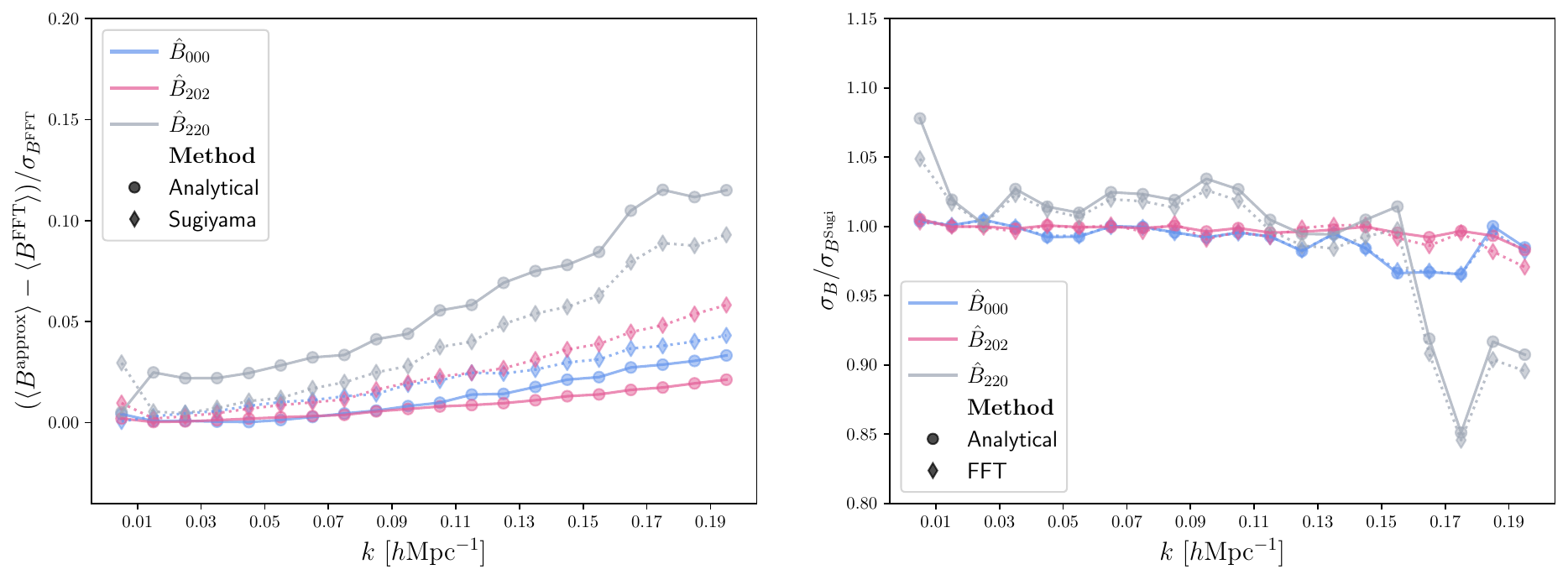}
\caption{\label{fig:shot_comparison}
Left: Difference between the two approximate methods and the reference method, divided by the standard deviation of the bispectrum multipoles evaluated using the FFT method. Different colors represent different multipoles, and the two methods are distinguished by distinct markers.
Right: Comparison of the full bispectrum standard deviations obtained from the three methods.
}
\end{center}
\end{figure*}

To quantitatively assess the bias of both approximate methods relative to the FFT reference, we turn to Figure~\ref{fig:shot_comparison}. The left panel shows the normalized bias $(\langle B^{\rm approx} \rangle - \langle B^{\rm FFT} \rangle) / \sigma_{B^{\rm FFT}}$ for the analytical and Sugiyama methods, where $\sigma_{B^{\rm FFT}}$ denotes the $1\sigma$ uncertainty of the bispectrum multipoles, with the shot noise contribution estimated via the FFT method. For $B_{000}$ and $B_{202}$, both methods perform excellently, with normalized biases below $0.05\sigma$ across all scales. Notably, the analytical method achieves even smaller bias than the Sugiyama method for these configurations. For $B_{220}$, however, the situation differs: both methods exhibit biases around $0.1\sigma$ at small scales, where the Sugiyama method performs slightly better than the analytical method. The right panel compares the standard deviation ratio ${\rm \sigma}_B/{\rm \sigma}_{B^{\rm Sugi}}$, i.e., the standard deviation of the analytical and FFT methods normalized by that of the Sugiyama method. For $B_{000}$ and $B_{202}$, the differences are quite small except on small scales. For $B_{220}$, however, both FFT and analytical methods show larger variance ratios at large scales, indicating that they perform worse than the Sugiyama method in this regime, while at small scales their performance is slightly better.

\begin{figure*}
\begin{center}
\includegraphics[width=0.6 \textwidth]{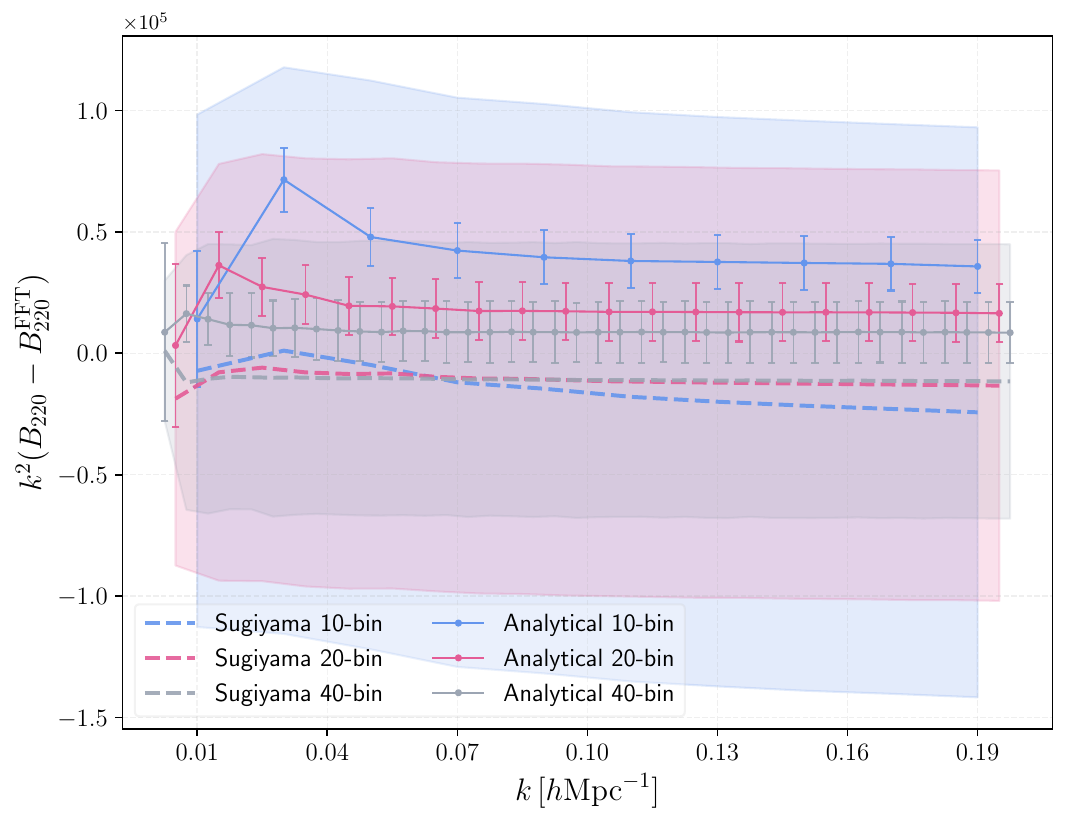}
\caption{\label{fig:residuals_comparison}
Residuals of the multipole $B_{220}$ when the shot noise is computed using the Sugiyama/analytical method, relative to the FFT method, measured from 200 MD-patchy mock catalogs. The dashed lines, along with the corresponding coloured shaded regions, represent the mean residual and the standard deviation for the Sugiyama method, while the solid lines with error bars represent the mean residual and the standard deviation for the analytical method. Different colours indicate different numbers of bins.}
\end{center}
\end{figure*}

Having established the bias characteristics at fixed binning (20 bins), we next investigate how bin width affects the performance of both approximate methods. Figure~\ref{fig:residuals_comparison} shows the mean and variance of the residuals for both methods relative to the FFT method, measured with 10, 20, and 40 $k$-bins. Several important patterns emerge. The analytical method for $B_{220}$ possesses remarkably small variance across all bin numbers, indicating high precision, but it consistently shows a clear positive offset in the residual mean. Conversely, the Sugiyama method exhibits smaller bias relative to the FFT method but suffers from much larger variance. As the number of bins increases (i.e., bin width decreases), the disadvantages of both methods are alleviated: the analytical method's bias diminishes, while the Sugiyama method's variance decreases. This bin-width dependence suggests that the analytical approximation becomes more accurate on finer $k$-grids, which is encouraging given that actual surveys often employ bin widths of $0.005\ h\text{Mpc}^{-1}$ or smaller. For practical applications of the analytical method for $B_{220}$ when dealing with thousands of mocks, we propose a calibration strategy: randomly select several hundred mocks, compute $B_{220}$ using both the FFT and analytical methods, and use the measured residual to calibrate the analytical estimates for the remaining mocks. On the other hand, when computational resources are not severely constrained, the FFT method remains preferable as our planned approach for future DESI analysis---despite nearly doubling the computational cost, it avoids the additional thin-shell approximation employed by the analytic and Rayleigh methods.


\section{CosmoNPC: an efficient toolkit for LSS clustering measurements}\label{sec:cosmonpc}

Based on the suite of acceleration techniques developed in this work, we have released \texttt{CosmoNPC}, an open-source Python/MPI package for large-scale structure clustering measurements.

\subsection{Design goals}

\texttt{CosmoNPC} aims to provide an efficient, user-friendly, and scalable tool for clustering statistics in both galaxy surveys and numerical simulations. The code fully incorporates the symmetry reductions, high-order power spectrum multipole accelerations, and analytical shot-noise treatment presented in this paper, while being engineered for practical data analysis workflows.

\subsection{Core functionality}

The current stable version supports:
\begin{itemize}
    \item \textbf{Power spectrum multipoles} (Yamamoto estimator), including auto and cross correlations; A \texttt{fast} mode implementing the modified high-order even multipole approximation described in \S{\ref{sec:level2}} is also available.
    \item \textbf{Bispectrum multipoles}, supporting multiple tracer combinations in Sugiyama estimator.
\end{itemize}

\subsection{Key features}

\begin{itemize}
    \item \textbf{Distributed parallelism:} Built on MPI, the code scales efficiently to multi-node computing environments, making it suitable for large simulations and survey data.
    \item \textbf{Pure Python implementation:} Written entirely in Python and built on top of the mature open-source scientific computing stack, \texttt{CosmoNPC} is easy to install, modify, and extend.
    \item \textbf{Native multitracer support:} Auto and cross correlations are supported from the ground up, facilitating multitracer cosmological analyses.
    \item \textbf{Flexible shot-noise handling:} For the Sugiyama estimator, both the analytical method derived in this work and an FFT-based numerical method are provided for cross-validation.
    \item \textbf{Flexible input formats:} The code accepts both periodic-box catalogs (Cartesian coordinates and peculiar velocities) and survey FITS files (containing RA, DEC, redshift, weights, etc.).
\end{itemize}

\subsection{Status and outlook}
The implementation of the estimators presented in this work is ongoing. Future releases will include the Scoccimarro bispectrum estimator, the new compressed bispectrum estimator introduced in this work, additional statistics, and further performance optimizations.

\section{Conclusion and discussion}\label{sec:conclusions}
In this paper, we focus on reducing the computational complexity of two- and three-point correlation estimators in Fourier space by minimizing the number of required FFTs and other expensive operations. Exploiting the fact that the multipole expansion results for the power spectrum and bispectrum are purely real or imaginary, we reduce the required number of sub-configurations by nearly half, achieving a twofold speedup. This reduction applies to virtually all estimators for two- and three-point multipole expansions. For certain special configurations of the Sugiyama estimator, additional symmetries can be exploited, leading to an overall speedup of up to a factor of four.

For the power spectrum, we further validate the feasibility of expanding high-order multipoles in terms of low-order ones. Although it has long been recognized that this approach can substantially reduce the number of required FFTs, its reliability has been questioned. For the first time, we verify the reliability of this approximation at the level of cosmological parameter constraints via full-shape fits, using survey data on the scale of DESI DR2 (though we employ BOSS data and enlarge its volume to match that of DESI). The results are consistent with the standard approach. For the bispectrum, inspired by two existing estimators, we propose a new estimator also based on the Tripolar basis. Specific multipoles such as $\hat{\mathcal{B}}_{\ell 0 \ell}$ in this estimator successfully reduce the number of sub-configurations that require extensive inverse FFT operations, achieving a speedup of nearly \((2\ell+1)\) times compared to the Sugiyama estimator while retaining nearly equivalent constraining power.

We present the analytic forms of the shot noise for the three bispectrum estimators discussed in this work, demonstrating that all of these shot noise terms can be computed with a computational complexity comparable to that of the power spectrum. For the Scoccimarro estimator, we successfully reproduce the form given in the original work. For the Sugiyama estimator, we find that, compared to the original method, the new analytic form exhibits reduced scatter and smaller deviations from the FFT-based reference for most multipoles. Moreover, its computational cost is negligible compared with the other two methods, yielding an overall speedup of nearly a factor of two for the entire estimator relative to the approach that computes the shot-noise contribution via FFT. The shot noise of our new estimator is directly derived from its analytic form. Throughout this process, we abstract the underlying mathematical physics integrals and obtain rather interesting results, extending our understanding of integrals of spherical harmonics over arbitrary regions in three-dimensional space. In particular, the familiar full-sphere average of spherical harmonics emerges as merely a special case of our derived results.

All of these techniques have been integrated into our publicly released package \texttt{CosmoNPC}.

A number of open questions merit further investigation, which we leave to future studies. Owing to computational limitations, most of our analysis has been restricted to the diagonal elements of the two-dimensional Sugiyama estimator, inevitably leading to incomplete information. In addition, the impact of both our new compressed estimator and the choice of shot-noise subtraction method on cosmological parameter constraints remains to be quantified. It would also be valuable to explore whether the analytical shot-noise method can be calibrated without relying on the expensive FFT-based reference---for instance, by deriving higher-order corrections to the thin-shell approximation to further reduce any residual bias. We intend to address these questions in future work.

\begin{acknowledgements}
We thank Naonori S. Sugiyama, Mike Shengbo Wang, Regina Demina, Héctor Gil-Marín and Jiamin Hou for helpful discussions. YX, RZ, GG, XW, YW and GBZ are supported by the National Key R \& D Program of China (2023YFA1607800, 2023YFA1607803), NSFC grants 11925303 and 11890691, and by the CAS Project for Young Scientists in Basic Research (No. YSBR-092). YX is also supported by the Chinese Scholarship Council (CSC) and the University of Edinburgh. RZ is supported by the CSC and the University of Portsmouth. YW is also supported by NSFC Grants (12273048, 12422301), and by National Key R\&D Program of China No. 2022YFF0503404. GBZ is also supported by science research grants from the China Manned Space Project with No. CMS-CSST-2021-B01, and the New Cornerstone Science Foundation through the XPLORER prize. FB is a University Research Fellow.
\end{acknowledgements}

\appendix

\section{Additional figures}\label{app:figures}
\begin{figure}[!h]
    \centering
    \includegraphics[width=0.8\textwidth]{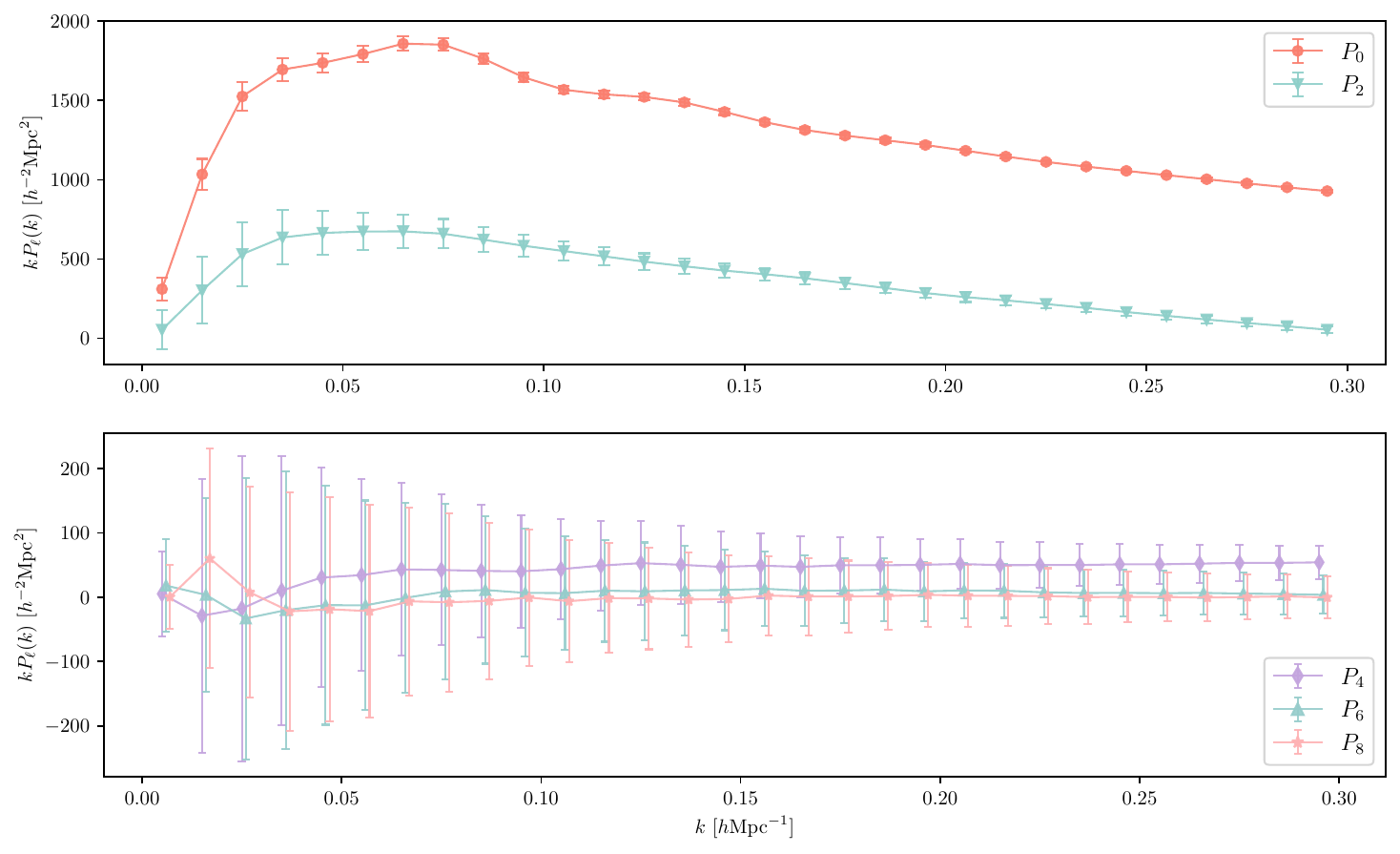}
    \caption{Mean and standard deviation of the traditional power spectrum multipoles measured from 2048 CMASS NGC MultiDark-Patchy mock catalogs in the redshift slice $z \in [0.2,0.5]$. The top and bottom columns show the $(\ell=0,2)$ and $(\ell=4,6,8)$ multipoles, respectively.}
    \label{fig:P_ell_app}
\end{figure}

Figure~\ref{fig:P_ell_app} presents the mean and standard deviation of the traditional power spectrum multipoles $P_0$, $P_2$, $P_4$, $P_6$, and $P_8$, measured from 2048 MultiDark-Patchy mock catalogs of the BOSS CMASS NGC sample in the redshift slice $z \in [0.2,0.5]$. These measurements serve as a baseline for validating the modified estimators $\widehat{P}_{\ell b}$ introduced in Section~\ref{sec:level2_validation}.

\begin{figure}[!h]
    \centering
    \includegraphics[width=\textwidth]{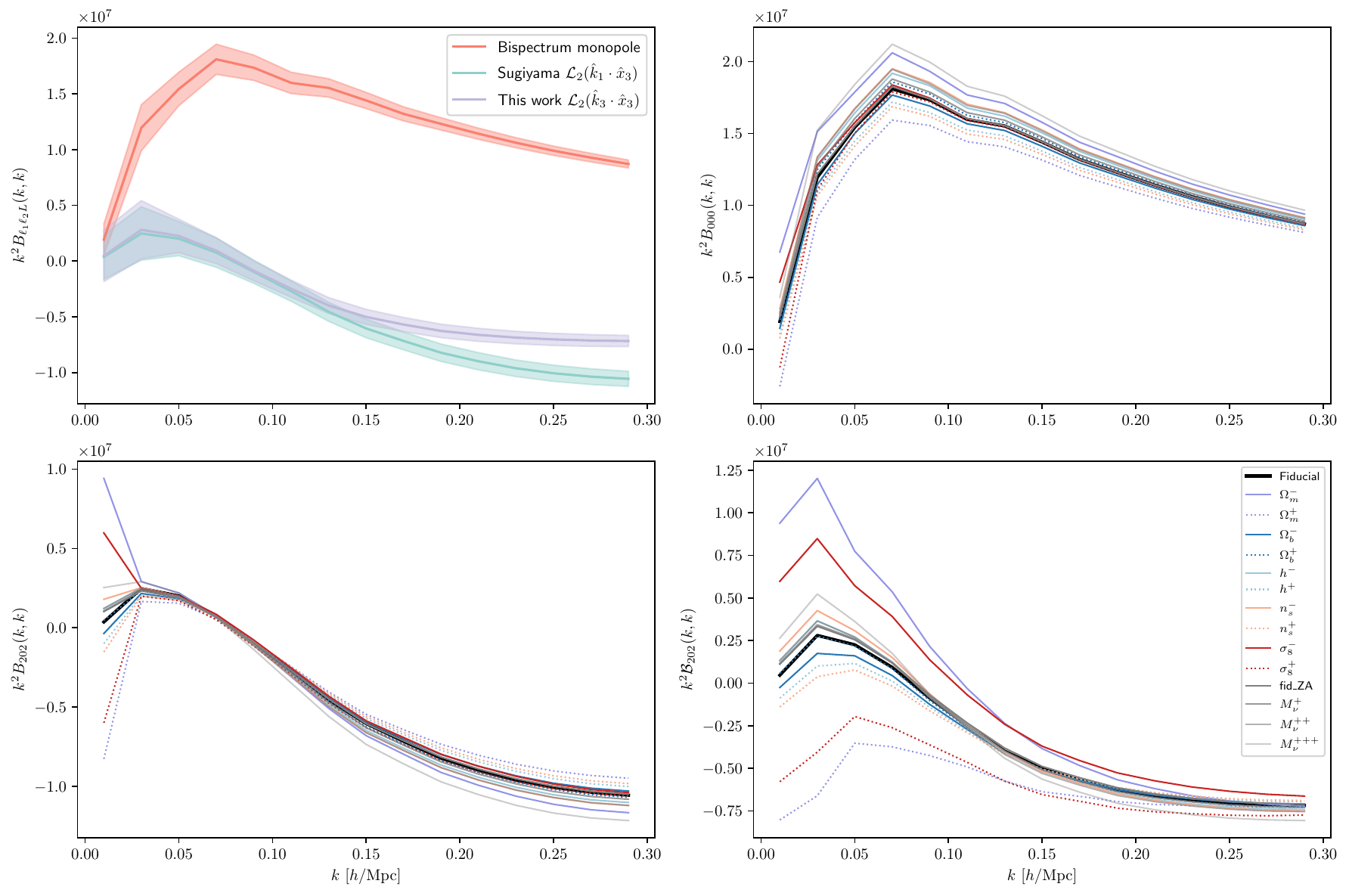}
    \caption{Top left: Mean (solid line) and standard deviation (color band) of the compressed bispectrum monopole, along with quadrupoles derived from two distinct compression directions, for fiducial-cosmology mocks. Top right: Mean shape of the measured bispectrum monopole as a function of varying cosmological parameters. Bottom left: Same as top right, but for the Sugiyama quadrupole. Bottom right: Same as top right, but for the quadrupole from this work.}
    \label{fig:B_multipole}
\end{figure}

Figure~\ref{fig:B_multipole} shows the bispectrum monopole and quadrupoles measured from the Molino mocks, illustrating the shape and cosmological dependence of the new estimator $\widehat{\mathcal{B}}_{202}$ compared to the Sugiyama quadrupole $\widehat{B}_{202}$.

\section{Identities of spherical harmonic functions}{\label{sec:ylm}}
It is straightforward to verify that the commonly used (complex) spherical harmonics,
\begin{equation}\label{eq:def_sph}
Y_\ell^m (\theta,\phi) = (-1)^m \sqrt{\frac{2 \ell+1}{4 \pi} \frac{(\ell-|m|)!}{(\ell +|m|)!}} \, \mathcal{L}_\ell^{m}(\cos{\theta}) \, {\rm e}^{{\mathrm i}m \phi}\,,
\end{equation}
where $\mathcal{L}_\ell^{m}(\cos{\theta})$ are the associated Legendre polynomials, 
\begin{equation}
\begin{gathered}
        \mathcal{L}_{\ell}^m(x)=(-1)^m\left(1-x^2\right)^{m / 2} \frac{{\rm d}^m}{{\rm d} x^m}\mathcal{L}_{\ell}(x)\ ,\quad (m\ge 0)\,,\\
        \mathcal{L}_{\ell}^{-m}(x)=(-1)^m \frac{(\ell-m)!}{(\ell+m)!} \mathcal{L}_{\ell}^m(x) \ ,\quad (m< 0)\,,
\end{gathered}
\end{equation}
satisfy the following relations~\cite{arfken2011mathematical}:
\begin{equation}\label{eq:sph_prop}
\begin{gathered}
Y_{\ell}^{m}(-\hat{r}) = (-1)^\ell Y_{\ell}^{m}(\hat{r})\,, \\
Y_{\ell}^{m}(\hat{r}) = (-1)^m Y_{\ell}^{-m*}(\hat{r})\,.
\end{gathered}
\end{equation}
Combining these two equations yields
\begin{equation}\label{eq:sphiv}
   Y_{\ell}^{-m}(\hat{r})= (-1)^{\ell+m}Y_{\ell}^{m*}(-\hat{r})\,.
\end{equation}
For reference, we list the values of several special spherical harmonics:
\begin{equation}
\begin{gathered}
    Y_{\ell}^m(\hat{z}) = Y_{\ell}^m(0,\phi)=\sqrt{\frac{2\ell +1}{4\pi}}\delta_{m0}^{\rm K}\,,\\
    Y_{\ell}^0(\theta,\phi) =\sqrt{\frac{2\ell +1}{4\pi}} \mathcal{L}_\ell(\cos{\theta})\,.
\end{gathered}
\end{equation}

The spherical harmonics are orthonormal,
\begin{equation}\label{eq:ylm_orthonormal}
    \int \mathrm{d}^2 \hat{r} \, Y_{\ell}^{m}(\hat{r})Y_{\ell'}^{m'*}(\hat{r}) = \delta^{\rm K}_{\ell \ell'}\delta^{\rm K}_{mm'}\,,
\end{equation}
from which it immediately follows that
\begin{equation}\label{eq:single_ylm_integral}
    \int \mathrm{d}^2 \hat{r} \, Y_{\ell}^{m}(\hat{r}) \propto \delta^{\rm K}_{\ell 0}\delta^{\rm K}_{m0}\,.
\end{equation}

Real-form spherical harmonics $X_{\ell, m}$ are obtained by a simple linear combination of the standard spherical harmonics,
\begin{equation}\label{real_sph}
X_{\ell m}= \begin{cases}\frac{{\mathrm i}}{\sqrt{2}}\left(Y_{\ell}^m-(-1)^m Y_{\ell}^{-m}\right), &m<0 \\ Y_{\ell}^0, & m=0 \\ \frac{1}{\sqrt{2}}\left(Y_{\ell}^{-m}+(-1)^m Y_{\ell}^m\right), & m>0 ,\end{cases}
\end{equation}
and satisfy the parity relation
\begin{equation}
    X_{\ell m}(\hat{r}) = (-1)^\ell X_{\ell m}(-\hat{r})\,,
\end{equation}
while the symmetries between $X_{\ell \pm m}$ no longer hold.

In this work, the Racah-normalized complex and real spherical harmonic functions are defined as $y_{\ell}^{m}(\hat{r}) =\sqrt{{4\pi}/\left(2\ell+1\right)}Y_{\ell}^{m}(\hat{r})$ and $x_{\ell m}(\hat{r}) = \sqrt{{4\pi}/\left(2\ell+1\right)}X_{\ell m}(\hat{r})$, respectively. One can easily write
\begin{equation}
y_{\ell_1}^{n}(\theta_{Q'},\phi_{Q'})y_{\ell_2}^{-n}(\theta_{R'},\phi_{R'}) 
=  (-1)^n \mathcal{L}_{\ell_1}^n(\cos \theta_{Q'}) \mathcal{L}_{\ell_2}^{-n}(\cos \theta_{R'})\,,
\end{equation}
where the condition $|\phi_{Q'}-\phi_{R'}|=\pi$ introduces an additional phase ${\rm{e}}^{{\mathrm i}n(\phi_{Q'}-\phi_{R'})}=(-1)^n$.

The addition theorem of spherical harmonics states that
\begin{equation}\label{eq:add_theorem}
    \mathcal{L}_{\ell}(\hat{r}_1 \cdot\hat{r}_2 )=\frac{4\pi}{2\ell+1}\sum_m Y_\ell^{m*}(\hat{r}_1)Y_\ell^{m}(\hat{r}_2)\,.
\end{equation}
Utilizing Eq.~\eqref{eq:def_sph}, we obtain
\begin{equation}
\begin{aligned}
\mathcal{L}_\ell \left(\hat{r}_1 \cdot\hat{r}_2 \right)
&=\mathcal{L}_\ell \left(\cos \theta_1  \right)\mathcal{L}_\ell \left(\cos \theta_2  \right) \\
&\quad +2\sum_{m=1}^\ell \frac{\left(\ell-m\right)!}{\left(\ell+m\right)!} \mathcal{L}_\ell^{m}\left(\cos \theta_1\right) \mathcal{L}_\ell^m\left(\cos \theta_2\right) \cos m (\phi_1-\phi_2)\,.
\end{aligned}
\end{equation}
Here we have chosen the $z$-axis to lie along $-(\bs{r}_1 +\bs{r}_2 )/|\bs{r}_1 +\bs{r}_2 |$, and $\theta_i$, $\phi_i$ denote the polar and azimuthal angles of $\bs{r}_i$ in this spherical coordinate system. In this frame we also have $|\phi_1-\phi_2| = \pi$, which allows us to simplify the expression to
\begin{equation}\label{eq:add_associated_L}
\begin{aligned}
\mathcal{L}_\ell \left(\hat{r}_1 \cdot\hat{r}_2 \right)  = \sum_{m=-\ell}^{\ell} \mathcal{L}_{\ell}^m\left(\cos \theta_1\right) \mathcal{L}_{\ell}^{-m}\left(\cos \theta_2\right)\,.
\end{aligned}
\end{equation}

Spherical harmonics can also be used to expand plane waves via the Rayleigh expansion,
\begin{equation}\label{eq:sph_wave_expan}
    {\rm e}^{{\mathrm i} \bs{k} \cdot \bs{r}}=4 \pi \sum_{\ell=0}^{\infty} \sum_{m=-\ell}^\ell {\mathrm i}^\ell j_\ell(k r) Y_\ell^{m*}(\hat{k})Y_\ell^m\left(\hat{r}\right)\,.
\end{equation}
Combining this with Eq.~\eqref{eq:ylm_orthonormal}, we obtain
\begin{equation}\label{eq:sugi_ylm}
    \begin{aligned}
        \int \frac{\mathrm{d}^2 \hat{k}}{4 \pi} Y_{L}^{M}(\hat{k})\mathrm{e}^{{\mathrm i} \bs{k} \cdot \bs{r}}
        &=  {\mathrm i}^L j_L (k r) Y_{L}^{M}(\hat{r})\,.
    \end{aligned}
\end{equation}

\section{Definition and properties of Wigner $D$-functions}\label{sec:wignerD}
Let $\tilde{H}_{\ell} \equiv \operatorname{Span}\bigl(\{\ket{\ell m}\mid m=-\ell,-\ell+1,\dots,\ell\}\bigr)$ be the representation space spanned by the angular momentum eigenstates. Under a rotation $\hat{\mathcal{R}}$, a state transforms as
\begin{equation}
    \ket{\ell m}' = \hat{\mathcal{R}}\ket{\ell m}\,.
\end{equation}
Inserting a resolution of identity gives
\begin{equation}
    \hat{\mathcal{R}}\ket{\ell m} = \sum_{\ell' n} \ket{\ell' n}\bra{\ell' n}\hat{\mathcal{R}}\ket{\ell m}\,.
\end{equation}
Since $\tilde{H}_{\ell}$ is irreducible, it is closed under rotations: $\hat{\mathcal{R}}\ket{\ell m}\in\tilde{H}_{\ell}$. Using the orthonormality $\braket{\ell m|\ell' m'} = \delta_{\ell\ell'}\delta_{mm'}$, we have $\braket{\ell m|\hat{\mathcal{R}}|\ell' m'} \propto \delta_{\ell\ell'}$. Consequently,
\begin{equation}
\begin{aligned}
\hat{\mathcal{R}}\ket{\ell m}
&= \sum_n \ket{\ell n}\bra{\ell n}\hat{\mathcal{R}}\ket{\ell m} \\
&\equiv \sum_n D_{nm}^{\ell}(\hat{\mathcal{R}})\ket{\ell n}\,,
\end{aligned}
\end{equation}
where $D_{nm}^{\ell}(\hat{\mathcal{R}})$ is the Wigner $D$-matrix, the irreducible representation matrix of $\mathrm{SO}(3)$ on $\tilde{H}_{\ell}$. Parametrizing the rotation by Euler angles $(\alpha,\beta,\gamma)$, the Wigner $D$-functions take the form \cite{Varshalovich:1988ifq}
\begin{equation}\label{eq:D_euler}
    D_{nm}^{\ell}(\hat{\mathcal{R}}(\alpha,\beta,\gamma)) = {\rm e}^{-{\rm i} n\alpha}\, d_{nm}^{\ell}(\beta)\, {\rm e}^{-{\rm i} m\gamma}\,,
\end{equation}
with $d_{nm}^{\ell}(\beta)$ the Wigner $d$-function.

We now list some useful properties. Unitarity follows from
\begin{equation}\label{eq:uni_d}
\begin{aligned}
D_{nm}^{\ell}(\hat{\mathcal{R}}^{-1}) 
= \bra{\ell n}\hat{\mathcal{R}}^{-1}\ket{\ell m}
= \bra{\ell n}\hat{\mathcal{R}}^{\dagger}\ket{\ell m}
= D_{mn}^{\ell*}(\hat{\mathcal{R}})\,,
\end{aligned}
\end{equation}
where we used the unitarity of $\hat{\mathcal{R}}$ and the definition of the Hermitian conjugate. Spherical harmonics and Legendre functions appear as special cases:
\begin{equation}
\begin{gathered}
D_{m0}^{\ell*}(\alpha,\beta,\gamma)=\sqrt{\frac{4\pi}{2\ell+1}}\, Y_\ell^{m}(\beta,\alpha)\,, \\[2mm]
D_{00}^{\ell}(\alpha,\beta,\gamma) = d_{00}^{\ell}(\beta) = \sqrt{\frac{4\pi}{2\ell+1}}\, Y_\ell^{0}(\beta,\alpha) = \mathcal{L}_{\ell}(\cos\beta)\,.
\end{gathered}
\end{equation}

Next, we derive the coupling rule (Clebsch–Gordan series) for $D$-functions. Expand the direct product state $\ket{\ell_1 n_1 \ell_2 n_2} \equiv \ket{\ell_1 n_1}\otimes\ket{\ell_2 n_2}$ in the coupled basis $\ket{JN}$:
\begin{equation}
\ket{\ell_1 n_1 \ell_2 n_2}
= \sum_{JN} \ket{JN}\braket{JN|\ell_1 n_1 \ell_2 n_2}\,,
\end{equation}
where $\braket{JN|\ell_1 m_1 \ell_2 m_2}$ is the Clebsch–Gordan coefficient. Applying a rotation $\hat{\mathcal{R}}(\alpha,\beta,\gamma)$ to both sides yields
\begin{equation}
\begin{aligned}
&\sum_{n_1'n_2'} D^{\ell_1}_{n_1'n_1}(\alpha,\beta,\gamma)
D^{\ell_2}_{n_2'n_2}(\alpha,\beta,\gamma)\ket{\ell_1 n_1' \ell_2 n_2'} \\
&= \sum_{JNN'} D^{J}_{N'N}(\alpha,\beta,\gamma)\ket{JN'}\braket{JN|\ell_1 n_1 \ell_2 n_2}\,.
\end{aligned}
\end{equation}
Multiplying on the left by $\bra{\ell_1 m_1 \ell_2 m_2}$ and using its expansion in the coupled basis,
\begin{equation}
\bra{\ell_1 m_1 \ell_2 m_2} = \sum_{J'M'} \braket{\ell_1 m_1 \ell_2 m_2 | J'M'}\bra{J'M'}\,,
\end{equation}
together with the orthonormality of the bases and the replacement of Clebsch–Gordan coefficients by $3$-$j$ symbols, we finally obtain the Clebsch–Gordan series for $D$-functions:
\begin{equation}\label{eq:D_couple}
\begin{aligned}
&D^{\ell_1}_{m_1 n_1}(\alpha,\beta,\gamma) D^{\ell_2}_{m_2 n_2}(\alpha,\beta,\gamma) \\
&= \sum_{J=|\ell_1-\ell_2|}^{\ell_1+\ell_2}\sum_{M'N} (-1)^{M'+N}(2J+1)
\begin{pmatrix} \ell_1 & \ell_2 & J \\ m_1 & m_2 & -M' \end{pmatrix}
\begin{pmatrix} \ell_1 & \ell_2 & J \\ n_1 & n_2 & -N \end{pmatrix}
D^{J}_{M'N}(\alpha,\beta,\gamma)\,.
\end{aligned}
\end{equation}
This equation represents the decomposition of the direct product of irreducible representations of the rotation group. The coupling rules for spherical harmonics $Y_{\ell}^m$ follow as a special case.

\section{Spherical Harmonic Functions Under Coordinate Rotations}\label{sec:rot_Ylm}

The spherical harmonic $Y_{\ell}^{m}(\hat{x})$ represents the state vector $\ket{\ell m}$ in the spherical coordinate basis $\ket{\hat{x}}$:
\begin{equation}
    Y_{\ell}^{m}(\hat{x}) \equiv \braket{\hat{x}|\ell m}\,.
\end{equation}

Applying the representation $\ket{\hat{x}}$ to both sides of Eq.~(\ref{eq:rot_state}) yields
\begin{equation}
    \braket{\hat{x}|\hat{\mathcal{R}}|\ell m} = \braket{\hat{\mathcal{R}}^{\dagger}\hat{x}|\ell m} = \braket{\hat{\mathcal{R}}^{-1}\hat{x}|\ell m} = Y_{\ell}^{m}(\hat{\mathcal{R}}^{-1}\hat{x})\,,
\end{equation}
and
\begin{equation}
    \bra{\hat{x}}\sum_n D_{nm}^{\ell}(\hat{\mathcal{R}})\ket{\ell n} = \sum_n D_{nm}^{\ell}(\hat{\mathcal{R}})\, Y_{\ell}^{n}(\hat{x})\,.
\end{equation}

Combining the last two equations and using the unitarity of the $D$-functions given by Eq. (\ref{eq:uni_d}), we obtain
\begin{equation}
    Y_{\ell}^m(\hat{\mathcal{R}}\hat{x}) = \sum_n D_{mn}^{\ell*}(\hat{\mathcal{R}})\, Y_{\ell}^n(\hat{x})\,.
\end{equation}

\section{Proofs of key steps in lossless acceleration}{\label{sec:proofs}}
\subsection[Proof of Eq. (Key 1)]{Proof of Eq.~\eqref{eq:key1}}
Taking the space inversion and complex conjugation of $S_{\ell}^{m}(\bs{k})$ gives $S_{\ell}^{m*}(-\bs{k})=\int {\rm d}^3x \,R(\bs{x})Y_{\ell}^{m*}(\hat{x})\mathrm{e}^{-\mathrm{i}\bs{k}\cdot\bs{x}}$. Recalling the second line of Eq.~\eqref{eq:sph_prop}, we obtain
\begin{equation}
\begin{aligned}
S_{\ell}^{m*}(-\bs{k}) 
&= (-1)^m \int {\rm d}^3x \,R(\bs{x})Y_{\ell}^{-m}(\hat{x})\mathrm{e}^{-\mathrm{i}\bs{k}\cdot\bs{x}}\\   
& = (-1)^m S_{\ell}^{-m}(\bs{k})\,.
\end{aligned}
\end{equation}

\subsection[Proof of Eqs. (F lm SIC and mathcalF)]{Proof of Eqs.~\eqref{eq:F_lm_SIC} and~\eqref{eq:mathcalF}}
Combining Eq.~\eqref{eq:key1} and Eq.~\eqref{eq:sphiv} yields Eq.~\eqref{eq:F_lm_SIC}. From this, we can further derive
\begin{equation}
    F_{\ell}(\bs{k})=(-1)^{\ell }F_{\ell}^*(-\bs{k})\,,
\end{equation}
and therefore Eq.~\eqref{eq:mathcalF}:
\begin{equation}
    \begin{aligned}
        \int \frac{\dd^2 \hat{k}}{4 \pi} F_0(\bs{k}) \left[\mathcal{G}_\ell^*(-\bs{k})\right]^*
        &=
        \int \frac{\dd^2 \hat{k}}{4 \pi} F_0^*(-\bs{k}) \mathcal{G}_\ell(\bs{-k})\\
        &=
        \left[\int \frac{\dd^2 \hat{k}}{4 \pi} F_0(\bs{k}) \mathcal{G}^*_\ell(\bs{k})\right]^*\,.
\end{aligned}
\end{equation}

\subsection[Proof of Eq. (SUGI F and G)]{Proof of Eq.~\eqref{eq:sugi_F_and_G}}
From Eq.~\eqref{eq:key1} we have $\left.\delta n_L^{-M}\right|_{\mathrm{FFT}}(\bs{k})=(-1)^{M}\left.\delta n_L^{M*}\right|_{\mathrm{FFT}}(-\bs{k})$. Substituting this into $G_L^{-M}(\bs{x})$ gives
\begin{equation}
\begin{aligned}
    G_L^{-M}(\bs{x}) 
    = (-1)^M \int \frac{{\rm d}^3 k}{(2 \pi)^3} \, \mathrm{e}^{\mathrm{i} \bs{k} \cdot \bs{x}} \frac{\left.\delta n_L^{M*}\right|_{\mathrm{FFT}}(-\bs{k})}{W_{\text {mass }}(\bs{k})}\,,
\end{aligned}
\end{equation}
where $W_{\text {mass }}(\bs{k})$ is real and isotropic, i.e., $W_{\text{mass }}(\bs{k})=W_{\text {mass }}(k)=W_{\text {mass }}^*(k)$. Changing the integration variable $\bs{k} \to -\bs{k}$, we obtain
\begin{equation}{\label{G_minus_M}}
\begin{aligned}
    G_L^{-M}(\bs{x}) 
    &= (-1)^M\int \frac{{\rm d}^3 k}{(2 \pi)^3} \, \mathrm{e}^{-\mathrm{i} \bs{k} \cdot \bs{x}} \frac{\left.\delta n_L^{M*}\right|_{\mathrm{FFT}}(\bs{k})}{W_{\text {mass }}(k)} \\
    &= (-1)^M G_L^{M*}(\bs{x})\,.
\end{aligned}
\end{equation}\par
Similarly,
\begin{equation}
\begin{aligned}
F_{\ell}^{-m}(\bs{x} ; k) 
& =\int \frac{{\rm d}^2 \hat{k}}{4 \pi}  \, \mathrm{e}^{\mathrm{i} \bs{k} \cdot \bs{x}} y_{\ell}^{-m *}(\hat{k}) \frac{\left.\delta n ^* \right|_{\mathrm{FFT}}(-\bs{k})}{W_{\text {mass }}(\bs{k})} \\
& = \frac{1}{N_{\text{mode}}(k)} \int \frac{{\rm d}^3 k}{(2 \pi)^3} \, \mathrm{e}^{\mathrm{i} \bs{k} \cdot \bs{x}} \Phi(k)  y_{\ell}^{-m *}(\hat{k}) \frac{\left.\delta n^*\right|_{\mathrm{FFT}}(-\bs{k})}{W_{\text {mass }}(k)}\,,
\end{aligned}
\end{equation}
where $\Phi(k)$ is the radial binning mask defined as
\begin{equation}\label{eq:binning_mask}
\Phi(k)=\left\{\begin{matrix}
 &1\,, &\quad {k}_{\rm c}-\Delta k /2 \le |\bs{k}| \le {k}_{\rm c}+\Delta k /2\,,\\
&0\,, & {\rm otherwise}\,.
\end{matrix}\right.
\end{equation}
Here $N_{\text{mode}}(k)$ is the number of mesh grids in a given $k$-shell centered at ${k}_{\rm c}$ with shell thickness $\Delta k$. Changing the integration variable $\bs{k} \to -\bs{k}$ and using Eq.~\eqref{eq:sphiv}, we find
\begin{equation}{\label{F_minus_m}}
\begin{aligned}
F_{\ell}^{-m}(\bs{x} ; k) 
&= \frac{(-1)^{\ell+m}}{N_{\text{mode}}(k)} \int \frac{{\rm d}^3 k}{(2 \pi)^3} \, \mathrm{e}^{-\mathrm{i} \bs{k} \cdot \bs{x}} \Phi(k)  y_{\ell}^{m }(\hat{k}) \frac{\left.\delta n^*\right|_{\mathrm{FFT}}(\bs{k})}{W_{\text {mass }}(k)}\\
&= (-1)^{\ell+m} F_{\ell}^{m*}(\bs{x} ; k)\,.
\end{aligned}
\end{equation}\par

\section{Real decomposition of bipolar and tripolar isotropic bases}{\label{sec:real_decomp}}
To fully take advantage of rFFT, we must decompose the bipolar and tripolar isotropic bases in terms of real spherical harmonics. The complex and real spherical harmonics are related by a unitary transformation,
\begin{equation}{\label{eq:unitary}}
\begin{aligned}
    Y_{\ell}^m(\hat{r})&=\sum_n U_{m n}^{\ell} X_{\ell n}(\hat{r})\,,\\
    X_{\ell m}(\hat{r})&=\sum_n V_{m n}^{\ell} Y_{\ell }^n(\hat{r})\,,
\end{aligned}
\end{equation}
where $\mathbf{U}^{\ell}$ and $\mathbf{V}^{\ell}$ are both unitary matrices; $\mathbf{V}^{\ell}$ is defined by Eq.~\eqref{real_sph} and $\mathbf{U}^{\ell}=(\mathbf{V}^{\ell})^{-1}$. By the definition of a unitary matrix, ${\mathbf{U}^{\ell}}^{\dagger}{\mathbf{U}^{\ell}}=1$, i.e.,
\begin{equation}
    \left({\mathbf{U}^{\ell}}^{\dagger}{\mathbf{U}^{\ell}}\right)_{nn'}
    =\sum_{m}\left({\mathbf{U}^{\ell}}^{\dagger}\right)_{nm}{\mathbf{U}^{\ell}}_{mn'}
    =\sum_{m}U_{mn}^{\ell * }U_{mn'}^{\ell}
    =\delta_{nn'}\,.
\end{equation}

For bipolar bases (Legendre polynomials), the spherical harmonic addition theorem~\eqref{eq:add_theorem} applies. Substituting the first line of Eq.~\eqref{eq:unitary} into the addition theorem, we obtain
\begin{equation}{\label{eq:real_addition}}
\begin{aligned}
\mathcal{L}_{\ell}(\hat{r}_1 \cdot\hat{r}_2 )
&=\frac{4\pi}{2\ell+1}\sum_{mnn'}U_{mn}^{\ell *}U_{mn'}^{\ell} X_{\ell n}(\hat{r}_1)X_{\ell n'}(\hat{r}_2)\\
&= \frac{4\pi}{2\ell+1}\sum_{n}X_{\ell n}(\hat{r}_1)X_{\ell n}(\hat{r}_2)\,,
\end{aligned}
\end{equation}
which is the addition theorem for real spherical harmonics. The similarity in mathematical form between the real and complex decompositions can be understood as follows: the Legendre polynomial can be written as a scalar product, $\mathcal{L}_{\ell}(\hat{r}_1 \cdot\hat{r}_2 )=4\pi/(2\ell+1)\,\mathbf{Y}_{\ell}^*(\hat{r}_1)\cdot \mathbf{Y}_{\ell}(\hat{r}_2)$, where $\mathbf{Y}_{\ell}= (Y_{\ell}^{-m},Y_{\ell}^{-m+1},\dots ,Y_{\ell}^{m})$, and scalar products are invariant under unitary transformations. For tripolar bases in which at least one subscript is zero---which reduce effectively to Legendre polynomials---Eq.~\eqref{eq:real_addition} remains valid.

In more general cases, however, the real decomposition differs substantially from the complex decomposition. This is because the tripolar basis involves three independent unitary transformations, $\mathbf{U}^{\ell_1}$, $\mathbf{U}^{\ell_2}$, and $\mathbf{U}^{L}$, one for each angular momentum index. In the bipolar case, the single unitary matrix $\mathbf{U}^{\ell}$ cancels out via ${\mathbf{U}^{\ell}}^{\dagger}\mathbf{U}^{\ell}=1$, leaving the form unchanged. In the tripolar case, however, the three unitary matrices are coupled through the Wigner $3$-$j$ symbol and cannot be eliminated by such a contraction, so the real decomposition does not reduce to the same simple form.

To obtain the correct real-form decomposition, we start from the complex decomposition of the isotropic tripolar basis $S_{\ell_1 \ell_2 L} (\hat{k}_1,\hat{k}_2,\hat{n})$ and substitute the unitary transformation into Eq.~\eqref{eq:sugi_base}, yielding
\begin{equation}
    \begin{aligned}
S_{\ell_1 \ell_2 L} (\hat{k}_1,\hat{k}_2,\hat{n})
&=   \sum_{n_1 n_2 N}\left [ \sum_{m_1 m_2M}\frac{U^{\ell_1}_{m_1 n_1}U^{\ell_2}_{m_2 n_2}U^{L}_{MN}}{H_{\ell_1 \ell_2 L}}\begin{pmatrix}
  \ell_1 &\ell_2& L \\ 
  m_1& m_2 & M
\end{pmatrix} \right]  x_{\ell_1 n_1}(\hat{k}_1)x_{\ell_2 n_2}(\hat{k}_2)x_{L N}(\hat{n})\\
&\equiv \sum_{n_1 n_2 N} \mathcal{A}^{n_1 n_2 N}_{\ell_1 \ell_2 L} x_{\ell_1 n_1}(\hat{k}_1)x_{\ell_2 n_2}(\hat{k}_2)x_{L N}(\hat{n})\,.
\end{aligned}
\end{equation}
The coefficients $\mathcal{A}^{n_1 n_2 N}_{\ell_1 \ell_2 L}$ can be pre-computed, and one must sum over $(2\ell_1+1)(2\ell_2+1)(2L+1)$ sub-configurations. Of course, many of these sub-configurations vanish upon explicit evaluation. For example, $S_{112}$, $S_{222}$, and $S_{224}$ contain 11, 25, and 37 non-zero sub-configurations, respectively. By comparison, the complex decomposition combined with our Level~1 acceleration yields only 5, 10, and 13 non-zero sub-configurations for the same bases.

\section{Details about the Space-Inversion algorithm}{\label{sec:space_inv}}
In Section \ref{sec:level1}, we propose a space-inversion algorithm to obtain $F_{\ell}(\bs{k})$ from $\mathcal{G}_{\ell}(\bs{k})$. The following is a single-process, demonstration-only implementation:\footnote{Indexes along one axis follow $0,1, \ldots ,N/2-1,-N/2,-N/2+1, \ldots ,-1$}:
\begin{lstlisting}
import numpy
def space_inversion(f):
    arr = numpy.empty_like(f)
    arr[0, 0, 0] = f[0, 0, 0]
    arr[0, 0, 1:] = numpy.flip(f[0, 0, 1:])
    arr[0, 1:, 0] = numpy.flip(f[0, 1:, 0])
    arr[1:, 0, 0] = numpy.flip(f[1:, 0, 0])
    arr[0, 1:, 1:] = numpy.flip(f[0, 1:, 1:])
    arr[1:, 0, 1:] = numpy.flip(f[1:, 0, 1:])
    arr[1:, 1:, 0] = numpy.flip(f[1:, 1:, 0])
    arr[1:, 1:, 1:] = numpy.flip(f[1:, 1:, 1:])
    return arr
\end{lstlisting}

A production MPI version would require significantly more complex handling of distributed memory and boundary conditions. Notably, it fails at grids of $F_{\ell}(\bs{k})$ corresponding to the Nyquist frequency; however, these grids contribute nothing to the evaluation of either the power spectrum or the bispectrum since we always apply an $k_{\rm max}$ cut well below the Nyquist frequency.

\section{Identities of Wigner 3-$j$ symbol}{\label{sec:3j}}
Wigner 3-$j$ symbols are an alternative to Clebsch–Gordan coefficients,
\begin{equation}
\left(\begin{array}{ccc}
j_1 & j_2 & j_3 \\
m_1 & m_2 & m_3
\end{array}\right) \equiv \frac{(-1)^{j_1-j_2-m_3}}{\sqrt{2 j_3+1}} \braket{j_1 m_1 j_2 m_2|j_3(-m_3)} \,.
\end{equation}

In this paper, we utilized one of its orthonormal relations,
\begin{equation}\label{eq:3j-orthonormal}
    \left(2 j_3+1\right) \sum_{m_1 m_2}\left(\begin{array}{ccc}
j_1 & j_2 & j_3 \\
m_1 & m_2 & m_3
\end{array}\right)\left(\begin{array}{ccc}
j_1 & j_2 & j_3^{\prime} \\
m_1 & m_2 & m_3^{\prime}
\end{array}\right)=\delta_{j_3, j_3^{\prime}}^{\rm K} \delta_{m_3, m_3^{\prime}}^{\rm K}\,,
\end{equation}
where we have assumed that quantum numbers already satisfy triangle conditions for non-zero values of  3-$j$ symbols.

\section{Effective integration volume}\label{sec:int_vol}
\begin{figure*}
\begin{center}
\includegraphics[width=0.5 \textwidth]{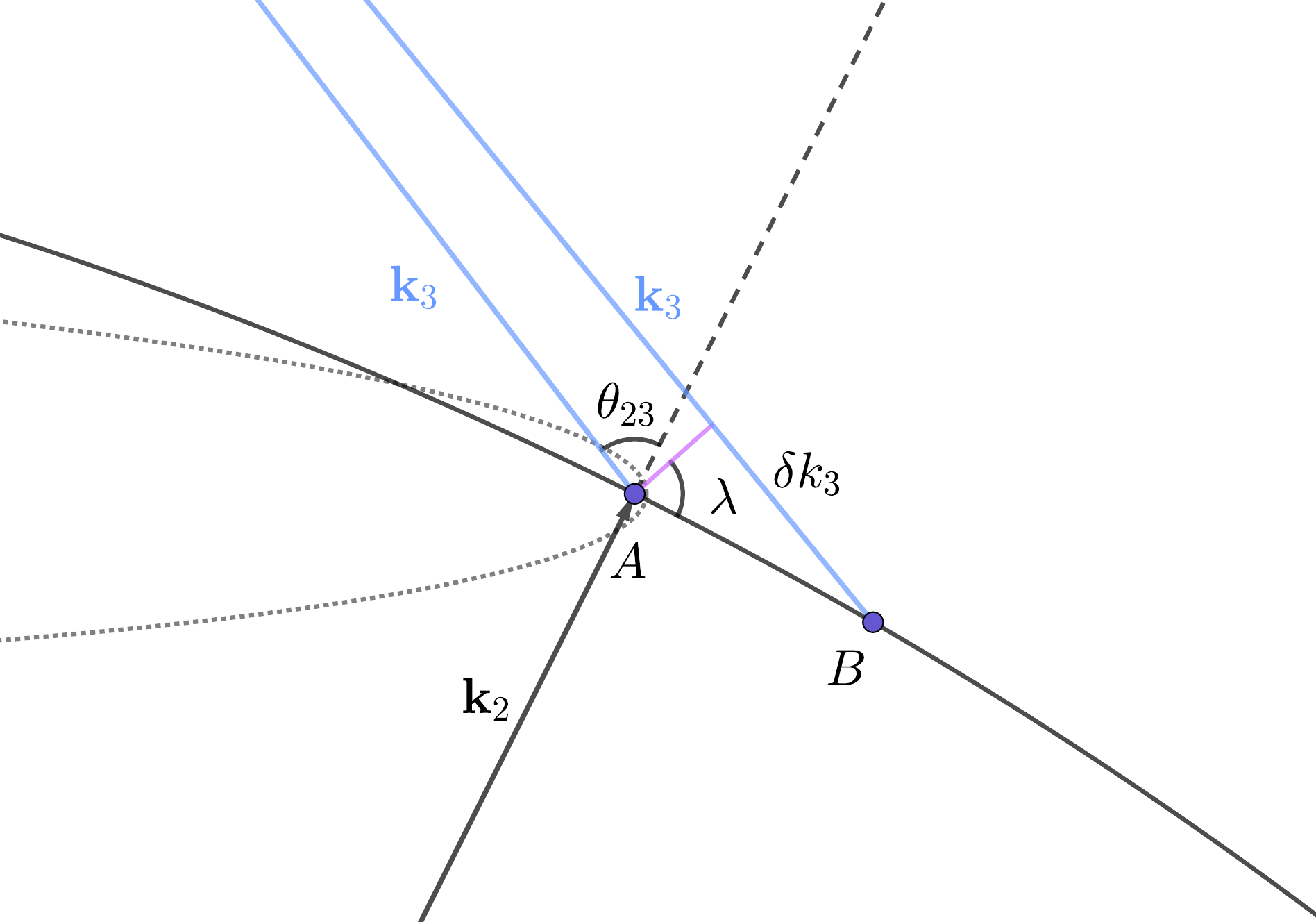}
\caption{\label{fig:V_dot}
A schematic diagram illustrating how to derive the total effective volume for a triangular configuration in Fourier space.
}
\end{center}
\end{figure*}

As shown in Figure \ref{fig:V_dot}, once $\bs{k}_1,\bs{k}_2$ are fixed, $\bs{k}_3$ only has a radial freedom $\delta k_3$, i.e., the origin of $\bs{k}_3$ can varies from $A$ to $B$ on the sphere with radius $k_2$. Since the purple line segment here is perpendicular to $\bs{k}_3$, we can quickly tell that $\theta_{23} = \lambda$, consequently, $\overline{AB}=\delta k_3 / \sin{\lambda} = \delta k_3 / \sin{\theta_{23}}$. Therefore 
\begin{equation}
    \begin{aligned}
        V_{\rm T}  &= 8 \pi^2 k_1^2 k_2\frac{\sin{\theta_{12}}}{\sin{\theta_{23}}} \delta k_1\delta k_2\delta k_3\\
        &=8 \pi^2 k_1 k_2 k_3 \delta k_1\delta k_2\delta k_3\,.
    \end{aligned}
\end{equation}
To obtain the second line in the equation above, we used the sine theorem of triangles, ${k_1}/{\sin \theta_{23}}={k_3}/{\sin \theta_{12}}$.

In Sugiyama estimator and our new estimator, $\bs{k}_3$ is compressed, thus the compressed total volume is 
\begin{equation}
    \begin{aligned}
        V_{\rm C} &\equiv \sum_{k_{3}}V_{\rm T}(k_3)\\
        &\simeq 8 \pi^2 k_1 k_2 \delta k_1\delta k_2 \int_{|k_1 - k_2|}^{k_1 + k_2} k_3 {\rm d } k_3\\
        &= 4\pi k_1^2 \delta k_1 \times  4\pi k_2^2 \delta k_2\,.
    \end{aligned}
\end{equation}

\section{Derivation of $g_\ell (t)$}\label{sec:g_ell}
Integrating and averaging over the spherical surface $S_P$ shown in the right panel of Figure~\ref{fig:sn_demo} and using Eq.~\eqref{eq:I_l0l}, we obtain
\begin{equation}
\int_{S_{P}} {\rm d}^2 S_p \,Y_\ell^{m}(-\hat{k}_3) 
=\int_0^{\pi} 2\pi k_2\sin\kappa \, k_2 \mathcal{L}_{\ell}(\cos\theta)Y_{\ell}^m(\hat{k}_1){\rm d}\kappa \,.
\end{equation}

When \(\theta \le \pi/2\), it is straightforward to show that
\[
\theta
=
\arctan\left(\frac{k_2\sin\kappa}{k_1+k_2\cos\kappa}\right)
=
\arctan\left(\frac{\sin\kappa}{t+\cos\kappa}\right)\,.
\]
Using the identity \(\cos[\arctan(x)]=1/\sqrt{1+x^2}\), we obtain
\begin{equation}
\cos\theta
=
\sqrt{
\frac{\cos^2\kappa+t^2+2t\cos\kappa}
{1+t^2+2t\cos\kappa}
}.
\end{equation}
For the obtuse branch, \(t+\cos\kappa<0\), this expression acquires an
additional negative sign. Therefore, for \(0\le t\le1\), the sign changes at
\(\kappa_0=\arccos(-t)\), which leads to Eq.~\eqref{eq:g_ell}. For \(t>1\),
\(t+\cos\kappa\) remains positive for all \(\kappa\in[0,\pi]\), and no split is
required.

When $t\gg 1$, the values of the spherical harmonics at different positions on the sphere are very close to their values along the symmetry axis $\bs{k}_1$. Consequently, $g_\ell (t)$ approaches unity in this limit. Conversely, as $t \to 0$, the sphere $S_P$ coincides with the sphere on which $\bs{k}_2$ lies. By Eq.~\eqref{eq:single_ylm_integral}, $g_\ell (t)$ approaches $0$ for all $\ell \neq 0$.

For the most common case $t=1$, trigonometric identities yield $1+\left[{\sin \kappa}/({1+\cos{\kappa}})\right]^2 =\cos^{-2}({\kappa}/{2})$. Since $\kappa \in [0,\pi]$, we have $\cos(\kappa/2) \ge 0$, and therefore $\cos{\theta} = \cos({\kappa}/{2})$. It follows that $g_{\ell}(1) = \frac{1}{2}\int_{0}^{\pi}{\rm d}\kappa \,\sin \kappa \, \mathcal{L}_{\ell}[\cos({\kappa}/{2})]$, which can be evaluated straightforwardly.

With the change of variables $x=\sqrt{1+t^2+2 t \cos \kappa}$, we also provide an equivalent but more practical form:
\begin{equation}
    g_\ell(t)=\frac{1}{2t} \int_{|t-1|}^{t+1} x\,\mathcal{L}_\ell\!\left(\frac{x^2+t^2-1}{2 x t}\right) {\rm d} x\,,\quad t\ge0\,.
\end{equation}


\allowdisplaybreaks

We list the analytical results for the first five orders of $g_{\ell}(t)$:
\begin{subequations}\label{eq:g_first_five}
\begin{align}
g_0(t) &= 1\,,\quad t\ge 0\,,
\label{eq:g0}
\\[4pt]
g_1(t) &=
\left\{
\begin{matrix}
 \dfrac{2t}{3}\,, & 0\le t<1\,,\\[4pt]
 1-\dfrac{1}{3t^2}\,, & t\ge 1\,,
\end{matrix}
\right.
\label{eq:g1}
\\[4pt]
g_2(t) &=
\left\{
\begin{matrix}
 0\,,& t=0\,,\\[4pt]
 \dfrac{1}{16 t^3}
 \left[
 2 t \left(5 t^2-3\right)
 +3 \left(t^2-1\right)^2
 \ln{\dfrac{t+1}{\left| t-1 \right|}}
 \right]\,, 
 & t\in(0,1)\cup (1,+\infty )\,,\\[4pt]
 \dfrac{1}{4}\,, & t=1\,,
\end{matrix}
\right.
\label{eq:g2}
\\[4pt]
g_3(t) &=
\left\{
\begin{matrix}
 0\,, & 0\le t<1\,,\\[4pt]
 \dfrac{(t^2-1)^2}{t^4}\,, & t\ge 1\,,
\end{matrix}
\right.
\label{eq:g3}
\\[4pt]
g_4(t) &=
\left\{
\begin{matrix}
 0\,,& t=0\,,\\[4pt]
 \dfrac{1}{192 t^5}
 \left[
 2 t \left(81 t^4-190 t^2+105\right)
 +15 \left(t^2-7\right) \left(t^2-1\right)^2
 \ln{\dfrac{t+1}{\left| t-1 \right|}}
 \right]\,, 
 & t\in(0,1)\cup (1,+\infty )\,,\\[4pt]
 -\dfrac{1}{24}\,, & t=1\,.
\end{matrix}
\right.
\label{eq:g4}
\end{align}
\end{subequations}

\bibliography{main}{}
\bibliographystyle{aasjournalv7}
\end{document}